\documentclass[12pt,preprint]{aastex}

\slugcomment{To appear in ApJ}

\shorttitle{The evolution of brown dwarf disks}
\shortauthors{Scholz et al.}

\begin{document}
\bibliographystyle{apj}


\title{Evolution of brown dwarf disks:\\ A Spitzer survey in Upper Scorpius}


\author{Alexander Scholz, Ray Jayawardhana}
\affil{Department of Astronomy \& Astrophysics, University of Toronto,
    60 St. George Street, Toronto, Ontario M5S3H8, Canada}
\email{aleks@astro.utoronto.ca}
\author{Kenneth Wood}
\affil{School of Physics \& Astronomy, University of St. Andrews, North Haugh, 
St. Andrews KY16 9SS}
\email{kw25@st-andrews.ac.uk}
\author{Gwendolyn Meeus}
\affil{Astrophysikalisches Institut Potsdam, An der Sternwarte 16, D-14482 Potsdam, 
Germany}
\email{gwen@aip.de}
\author{Beate Stelzer}
\affil{INAF - Osservatorio Astronomico di Palermo, Piazza del Parlamento 1, 
I-90134 Palermo, Italy}
\email{stelzer@astropa.unipa.it}
\author{Christina Walker, Mark O'Sullivan}
\affil{School of Physics \& Astronomy, University of St. Andrews, North Haugh, 
St. Andrews KY16 9SS}

\begin{abstract}
We have carried out a Spitzer survey for brown dwarf disks in the $\sim$5\,Myr old Upper Scorpius 
(UpSco) star forming region, using IRS spectroscopy from 8 to 12$\,\mu m$ and MIPS photometry at 
24$\,\mu m$. Our sample consists of 35 confirmed very low mass members of UpSco. Thirteen objects 
in this sample show clear excess flux at 24$\,\mu m$, explained by dust emission from a circum-sub-stellar 
disk. The spectral energy distributions (SEDs) of the remaining objects are consistent with 
pure photospheric emission. Objects without excess emission either have no disks at all or disks 
with inner opacity holes of at least $\sim 5$\,AU radii. Our disk frequency of $37\pm 9$\% is higher 
than what has been derived previously for K0-M5 stars in the same region (on a 1.8$\sigma$ confidence 
level), suggesting a mass-dependent disk lifetime in UpSco. The clear distinction between objects 
with and without disks as well as the lack of transition objects shows that disk dissipation inside 
5\,AU occurs rapidly, probably on timescales of $\lesssim 10^5$ years. For the objects with disks, 
most SEDs are uniformly flat with flux levels of a few mJy, well modeled as emission from dusty 
disks affected by dust settling to the midplane, which also provides indirect evidence for grain 
growth. The silicate feature around 10$\,\mu m$ is either absent or weak in our SEDs, arguing for a 
lack of hot, small dust grains. Compared with younger objects in Taurus, brown dwarf disks in UpSco 
show less flaring. By comparing SEDs of stars and brown dwarfs in UpSco, we find that dust settling 
is not a strong function of mass in this region. Taken together, these results clearly demonstrate 
that we see disks in an advanced evolutionary state: Dust settling and grain growth are ubiquituous 
in circum-sub-stellar disks at ages of 5\,Myr, arguing for planet forming processes in brown dwarf 
disks. For almost all our targets, results from high-resolution spectroscopy and high-spatial 
resolution imaging have been published before, thus providing a large sample of brown dwarfs for 
which information about disks, accretion, and binarity is available. We find that nine out of 13 
objects with disks do not accrete significantly. Hence, dusty disks can persist although the continuous 
accretion has stopped or dropped below measurable levels. Four objects with disks are binaries, three 
of them with (projected) separations $<30$\,AU. These objects likely harbour small disks truncated 
by the companion.
\end{abstract}



\section{Introduction}
\label{intro}

Young brown dwarfs are known to undergo a T Tauri phase similar to solar-mass
stars. A large fraction of them exhibits spectroscopic signatures of ongoing accretion and
outflows \citep[e.g.][]{2003ApJ...592..282J,2005ApJ...626..498M,2005ApJ...625..906M} combined 
with strong, T Tauri like variability \citep[e.g.][]{2005A&A...429.1007S,2006ApJ...638.1056S}. 
Moreover, near/mid infrared colour excess has been found, indicative of the presence of  
dusty circum-sub-stellar disks. It is this latter aspect of the T Tauri phase in brown 
dwarf evolution that we want to investigate in this paper.

Probing the properties and evolution of brown dwarf disks by analyzing their 
spectral energy distribution (SED) in the infrared and sub/mm regime is a useful 
test of brown dwarf origins, for example to distinguish between {\it in situ} formation
and formation as stellar embryos ejected from a multiple system by searching for
signatures of disk truncation \citep[see][for a summary]{2006astro.ph..2367W}. 
Additionally, disks are believed to be the nurseries of planets, and hence substellar 
disks allow us to study the early stages of planet formation, i.e. the process of grain 
and planetesimal growth, in an extreme case, constraining the efficiency and universality 
of current scenarios for the formation of planetary systems.
 
Until recently, our knowledge of substellar disks relied on a few case studies. In the
pre-Spitzer era, mid-infrared SEDs have been published for example by 
\citet{2002A&A...393..597N,2002ApJ...573L.115A,2004ApJ...609L..33M}, in total for $\sim 15$ 
brown dwarfs with ages of 1-2\,Myr, but only for one older object (the $\sim 8$\,Myr 
old 2M1207, \citet{2004A&A...427..245S}). By comparing the observed spectral energy 
distributions (SEDs) with model predictions, it has been found that young brown dwarfs 
appear to harbour disks with a diversity of geometries, which are interpreted as a consequence 
of disk evolution, consistent with findings for T Tauri stars. In some cases the opening angle 
of the disks increase with increasing distance from the star ("flared disks"), in others the 
disk is flat, which is explained as a consequence of grain growth and settling of the larger 
grains to the disk midplane \citep{2004A&A...426L..53A,2004A&A...427..245S}.

Larger samples have been investigated in the near-infrared, where ground-based observations 
are much easier \citep[e.g.][]{2001ApJ...558L..51M,2003AJ....126.1515J}, providing constraints 
on disk frequency as a function of age: Based on near-infrared (K-, L-band) colour excess, disk 
frequencies decline strongly between 1 and 5\,Myr, and only very few brown dwarfs retain their 
disks longer than that. Thus, the lifetimes for brown dwarf disks are in the range of 5\,Myr, 
which is comparable to the lifetimes of disks of solar-mass stars \citep{2001ApJ...553L.153H}. 

It is known, however, that the near-infrared ($1-4\,\mu m$) has major limitations for studies 
of brown dwarf disks \citep[e.g.][]{2000A&A...359..269C}. Problems in this wavelength regime
include, for example, the low contrast between disk and photosphere, the restriction to the 
innermost parts of the disks ($\lesssim 0.1\,AU$), and the presence of strong photospheric 
molecular bands for these cool objects. As a consequence, the disk census derived from 
near-infrared observations may be incomplete, and it is difficult to obtain reliable constraints 
on the disk lifetime. Mid-infrared observations give much better sensitivity for disk detections, 
because the photospheric flux rapidly declines at $\lambda >4\,\mu m$. The wavelength range 
$4-25\,\mu m$ is ideal to probe flaring geometry and inner disk hole sizes by comparing SED 
with model predictions. Moreover, an analysis of shape and intensity of the silicate feature 
around 10$\,\mu m$ is able to give detailed constraints on the warm disk chemistry 
\citep[e.g.][]{2003A&A...409L..25M}.

Spitzer provides unprecedented sensitivity between 3 and 24$\,\mu m$, and thus allows us for 
the first time to carry out efficient mid-infrared observations of large
object samples. Using Spitzer's IRS spectrograph, \citet{2005Sci...310..834A} find silicate
emission in six substellar disks at ages of 1-2\,Myr, and confirm previous claims for grain 
growth and crystallization. Spitzer also allows to push the limits in disk detection in terms 
of object age and mass \citep[e.g.][]{2006ApJ...639L..79R,2005ApJ...635L..93L}.

Only broad wavelength coverage and large object samples can overcome the large degree
of uncertainties when deriving disk properties, considering the numerous 
degeneracies inherent in radiation transfer modeling of dusty disks, and thus provide
reliable constraints on the disk parameters. This consideration is the main guideline
for our long-term program to characterize brown dwarf disks. In a first paper, we have 
combined 1.3\,mm measurements for 20 young brown dwarfs in Taurus with Spitzer IRAC/MIPS
photometry, to constrain fundamental parameters of substellar disks in a systematic way 
\citep{2006ApJ...645.1498S}. Our results, together with previous findings by 
\citet{2003ApJ...593L..57K}, show that brown dwarf disks can have masses of 
several Jupiter masses; at least 25\% have outer radii $> 10$\,AU. In Taurus there 
is no evidence for truncated disks due to an ejection process early in the life history 
of brown dwarfs, implying that most sub-stellar objects probably form in isolation.

In this second paper, we focus on evolutionary effects like disk dissipation, inner
disk clearing, dust settling, and grain growth, which are believed to be 
consequences or prerequisites of planet formation. The occurence of these processes in
brown dwarfs disks has already been confirmed, mostly at ages of 1-2\,Myr. The goal
of this paper is to probe disk evolution for a large sample of somewhat older objects.
The ideal target region for this project is the star forming association Upper Scorpius 
(UpSco): With an age of 5\,Myr and supposedly low age spread \citep[e.g.][]{1999AJ....117.2381P}, 
it is significantly older than the well-studied star forming regions in Chamaeleon, 
$\rho$\,Ophiuchus, and Taurus, the main targets in previous brown dwarf disk surveys, and 
thus presents disks at an evolved stage. Moreover, UpSco is nearby (distance $\sim 145$\,pc) 
and contains a large population of substellar objects, confirmed spectroscopically in recent
surveys \citep{2000AJ....120..479A,2004AJ....127..449M}. Small samples ($\le 10$) of brown dwarfs
in UpSco have already been studied in the near-/mid infrared by \citet{2003AJ....126.1515J} and
\citet{2006astro.ph..8395B}.

Using Spitzer IRS spectroscopy and MIPS 24$\,\mu m$ photometry, we carried out a disk survey 
for a large sample of 35 UpSco brown dwarfs. This dataset has been complemented by H$\alpha$ 
measurements and results from companions surveys (mostly from the literature), which gives us the 
opportunity to probe connections between disks, accretion, and binarity. In Sect. \ref{obs} we 
introduce the sample, describe the observations, the data reduction, and the extraction of SEDs. 
Subsequently, we give a general discussion of our SEDs, and derive an estimate of the disk frequency 
based on MIPS photometry (Sect. \ref{sed}). In the subsequent section, we compare the observed fluxes 
with models to constrain disk geometry (Sect. \ref{kenny}) and chemistry (Sect. \ref{gwen}). The
interpretation of these results is given in Sect. \ref{disc}, including a discussion of the
accretion/disk and disk/binarity connection. Finally, we give our conclusions in Sect. 
\ref{conc}.

\section{Targets, observations, data reduction}
\label{obs}

This paper is based on Spitzer observations of 35 very low mass objects in the
Upper Scorpius (UpSco) star forming region. The targets have been selected from 
the surveys of \citet{2000AJ....120..479A} and \citet{2004AJ....127..449M}, excluding 
objects later identified as non-members \citep{2004ApJ...609..854M}. 
From these spectroscopically confirmed young members of UpSco, we selected 35 objects 
with spectral type $\ge$M5, corresponding to effective temperatures $\lesssim$3000\,K 
\citep{2004ApJ...609..854M,2003ApJ...593.1093L} and thus masses $\lesssim 0.1\,M_{\odot}$ 
\citep{1998A&A...337..403B}. Although some
of these objects may have masses higher than the substellar limit at $0.08\,M_{\odot}$,
for the sake of simplicity we will call them brown dwarfs in the following. For a 
more detailed assessment of object masses see Sect. \ref{objprop}. Our target sample
thus comprises almost all brown dwarfs in UpSco known at the time of the observations
and is listed in Table \ref{objprop}. 
It does not include the very low mass objects in this region found recently by
\citet{2006astro.ph..2298S}. We note that the objects in our sample are mostly 
unbiased with respect to their disk environment, since they have been identified based 
on colour/magnitude diagrams and spectroscopic youth signatures.

A note on nomenclature: Objects from \citet{2000AJ....120..479A} are called 'UScoCTIO XXX' 
in the discovery paper, where XXX is an identification number; we will abbreviate them as 
uscoXXX in the following. The targets from \citet{2004AJ....127..449M} have originally been 
detected based on DENIS photometry and are thus called 'DENIS-P J' plus sky coordinates in 
DENIS (e.g., DENIS-P J155556.0-204518.5). For brevity, we will name these objects 'usd' 
followed by the first six digits of the sky coordinates, e.g. usd155556. To avoid
ambiguities, we include the coordinates of all sources in Table \ref{targets}.

All 35 objects were targets of Spitzer GO program \#20435 (PI: Ray Jayawardhana). Since 
six of these objects overlapped with the GTO program \#248 (PI: Subhanjoy Mohanty), they 
have been observed in the framework of the GTO campaign. For all objects we obtained IRS 
spectroscopy from 7.4 to 14.5$\mu m$ (module SL, 1st order) and MIPS 
images at 24$\mu m$. We aimed for a sensitivity of $\sim$0.2\,mJy, providing clear detections 
of the photospheric continuum at 10$\mu m$ even for our faintest targets and excellent 
upper limits at 24$\mu m$, allowing us a reliable distinction between objects with 
and without disks. The objects were observed with MIPS in one standard dither cycle 
with 10\,sec exposure time per position and small offsets, resulting in a total on-source 
time of $\sim 160$\,sec. The only exceptions were the six targets observed in program 
\#248, in these cases three cycles were carried out, in total $\sim 480$\,sec on-source 
time. IRS observations for all objects were carried out with two cycles, two nod positions, 
and 60\,sec ramp duration, resulting in a total on-source time of 240\,sec. The peak-up 
accuracy (i.e. the difference between pointing and requested position) was typically
0\farcs2 or less; most of this offset is along the slit and thus slit losses are negligible.

We used the post-BCD pipeline products for further reduction (pipeline version 
S13.2.0). Photometry on the MIPS post-BCD images was carried out using
{\it daophot} aperture photometry in IRAF with a large aperture of 13" and a sky annulus
with inner and outer radii of 20" and 32", respectively. After applying the aperture
correction given in the MIPS data handbook, version 3.2.1, fluxes were converted
to mJy. Fifteen brown dwarfs are clearly detected in the MIPS observations, with flux
levels between 0.35 and 10.7\,mJy. For the
non-detections, we estimated a generic upper limit from our images by measuring fluxes 
in a large number of regions containing only sky and no objects. These regions were
chosen to be close to the positions, where the objects are expected to be. After 
excluding some outliers, these measurements scatter between -0.2 and 0.2\,mJy. This 
is the flux level obtained from pure noise, and we thus infer a 2$\sigma$ upper limit 
of 0.4\,mJy for our MIPS non-detections. We note that this limit may vary somewhat from
object to object due to varying background level, but we prefer to use this statistically 
robust estimate instead of deriving upper limits for each object individually based on 
small number statistics. We estimate the uncertainty of our MIPS fluxes to be $\sim 10$\%, 
composed of uncertainties in calibration ($\sim 5$\%), aperture correction, and instrumental
magnitudes. (According to the suggestions in the MIPS data handbook, colour corrections
are negligible for our objects.) All 24$\mu m$ fluxes are listed in Table \ref{targets}.

For the IRS spectra, we started with the coadded spectra from the BCD data products,
pipeline version S13.2.0, and corrected for the sky background by subtracting the two nod
positions. Extraction of the spectra was done using the Spitzer IRS Custom Extractor 
{\it Spice}, version 1.3. All our targets are visible as profile peak along the slit, albeit 
some of them at the detection limit; flux levels at 10\,$\mu m$ are between 0.2 and 8.5\,mJy. 
Extraction was carried out with standard settings for the two nod positions, followed by 
averaging these two single datasets to produce the final spectrum for the respective object. 
Additionally, we extracted the sky noise at two positions in the slit not affected by the 
objects. After excluding outliers, these noise spectra provide us with an uncertainty 
estimate for each datapoint in the science spectra. Typical uncertainties of the individual
datapoints are in the range of 0.2-0.3\,mJy, where 7 spectra are affected of excessive noise 
in the range of $\sim 0.5$\,mJy, which can in most cases be attributed to background emission 
from the molecular cloud. Following the information in the IRS Data Handbook, version 2.0, we
expect additional flux calibration uncertainties of $\sim 10$\%. Because of low flux levels 
in most spectra and thus low signal-to-noise ratio, we binned all IRS spectra to a resolution
of $\sim 0.25\,\mu m$. This reduces the number of datapoints per spectrum from $\sim 100$ to
$\sim 20$ and improves the signal-to-noise ratio by a factor of $\sim 2$. All analysis using
the IRS data is done on these binned spectra (see Sect. \ref{gwen}).

All IRS spectra were truncated to the wavelength range between 7.5 and 12.5$\,\mu m$. Fluxes at 
longer wavelengths are increasingly affected by the '14 micron teardrop' effect, which is believed 
to be a type of scatter light, as described in the IRS data handbook. We calculated broadband fluxes 
at 9, 10, and 11$\,\mu m$ from the IRS spectra by averaging over a wavelength range of 1$\mu m$. 
Table \ref{targets} contains these values together with the average noise level for the individual
datapoints in the IRS spectrum for all objects. 

\begin{figure}
\includegraphics[angle=-90,width=13.8cm]{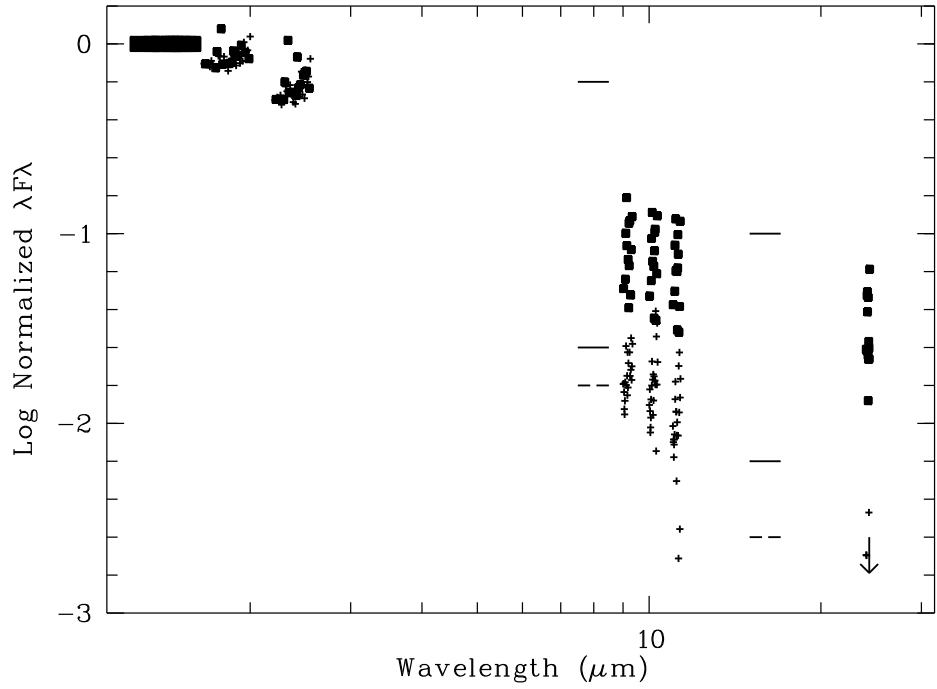}
\caption{ Broadband spectral energy distributions for all our sources, normalized to the 
J band flux. Objects with excess at 24\,$\mu m$ (see Fig. \ref{f1}) are shown as filled squares, 
objects without excess as plus symbols. The arrow at 24$\,\mu m$ gives a typical upper limit
for objects without MIPS detection. Dashed lines show average flux levels for K0-M5
stars without disks in UpSco, solid lines the typical range for stars with disks 
(taken from \citet{2006ApJ...651L..49C}). \label{f9}}
\end{figure}

Final spectral energy distributions (SEDs) were produced by combining IRS and MIPS fluxes
with near-infrared photometry from 2MASS and optical I-band photometry from 
\citet{2000AJ....120..479A} and \citet{2004AJ....127..449M}. After shifting 2MASS values 
to the CIT system \citep{2001AJ....121.2851C}, magnitudes were converted to fluxes using 
the zeropoints given by Skinner (1996), which are based on the CIT zeropoints in 
\citet{1976ApJ...208..390B}. Vice versa, we calculated $K - 24\,\mu m$ colours by 
converting the MIPS fluxes to magnitudes using the zeropoint of $7.14\pm 0.0815$\,Jy 
given in the MIPS data handbook. In Fig. \ref{f9} we show all SEDs for our sources,
normalized to the J-band fluxes. 

The brown dwarf usd160603 appears to have a faint neighbour in the MIPS image, at a
separation of 10". Since this object is not seen in 2MASS images, we assume that this
is a spurious source, most likely an asteroid. Because the separation is considerably larger
than the slitwidth of IRS (3.7"), the IRS spectrum of usd160603 is probably not affected 
by this neighbour. (Moreover, the spectrum has been taken three weeks before the MIPS image,
thus it is unlikely that the neighbour was at the same position, if it is indeed an asteroid.)
The large aperture of the MIPS photometry, however, does contain both objects. To correct 
for this contamination, we measured relative fluxes of both objects using a small aperture, 
and corrected the total flux accordingly. We obtain a flux of 2.84\,mJy for the brown dwarf 
and 1.50\,mJy for the (likely asteroidal) neighbour.

We aimed to include H$\alpha$ measurements for our targets in this paper. Of particular 
interest is the H$\alpha$ 10\% width, which is correlated with the disk accretion rate 
\citep{2004A&A...424..603N} and can therefore be used as a signature for accretion. For 19 
of our objects, H$\alpha$ 10\% widths have been published by \citet{2002ApJ...578L.141J}
or \citet{2005ApJ...626..498M}. We observed the remaining 16 objects using the high-resolution 
spectrographs UVES at ESO/VLT as part of the program 077.C-0323(B) and MIKE at the 6.5\,m Clay 
Magellan telescope on Las Campanas. For UVES we used dichroic 2 with central
wavelengths of 437 and 760\,nm and a 0\farcs8 slit, resulting in $R\sim 50000$. For MIKE we 
used a slit width of 0.7", $2 \times 2$ binning, and achieved a typical resolution of 
$R \sim 25000$. Data reduction and line measurements were carried out in a standard fashion,
see \citet{2006ApJ...638.1056S} for details. All H$\alpha$ 10\% widths are listed in Table 
\ref{targets}. Most of our targets have been observed recently with high spatial resolution 
to search for companions, either with HST \citep{2005ApJ...633..452K} or with adaptive optics 
\citep{2006A&A...451..177B}. We added the results from these surveys to Table \ref{targets}. 

\section{Spectral energy distributions}
\label{sed}

The aim of this section is to provide an overview of the SEDs for our targets and to
determine for how many objects we see evidence for excess mid-infrared emission due to 
disks. It turns out that based on a comparison of the Spitzer $24\mu m$ fluxes with 
photospheric fluxes expected for our targets, the disk frequency in our sample is 
$37\pm 9$\%. We will start with deriving fundamental parameters for the objects, 
to be able to estimate photospheric fluxes. Subsequently, we will discuss the SEDs and 
probe if disk excess is required to interpret the mid-infrared SEDs. 

\subsection{Fundamental parameters for the targets}
\label{objprop}

A reliable distance of Upper Scorpius has been determined based on the Hipparcos
measurements for the high-mass stars in this association. On average the
distance is 145\,pc \citep{1999AJ....117..354D}, with a spread of $\pm 15$\,pc, 
consistent with a roughly spherical shape for the association \citep{2002AJ....124..404P}. 

The age of the pre-main sequence population in Upper Scorpius has been subject of 
several recent studies. Since we are interested in evolutionary effects, a reliable
estimate of the age of UpSco is relevant. For the high-mass stars in this region, an 
age of 5-6\,Myr has been found from several different methods, with no evidence for 
age spread \citep{1989A&A...216...44D}. For the low-mass population, \citet{1999AJ....117.2381P} 
estimate an age of 5\,Myr based on an HR diagram, in agreement with the results for 
high-mass stars. According to their analysis, the age spread is not significant and 
probably smaller than 2\,Myr. This view has been challenged by \citet{2003A&A...404..913S}, 
who find an average age of 8-10\,Myr with a large spread of 1-20\,Myr for Upper Scorpius. 
Their HR diagram analysis, however, is mainly based on early-type stars, for which 1-20\,Myr 
isochrones quickly converge to the ZAMS and thus do not allow a clear discrimination 
between different ages. Therefore, we believe they overestimate the age spread.

Recently, \citet{2004ApJ...609..854M} measured gravities and temperatures from 
high-resolution spectroscopy for 11 of our targets. When compared with evolutionary 
tracks, 8 objects are clustered around the 5\,Myr isochrone, but three of them 
(usco104, 128, 130) appear to have ages $\lesssim 1$\,Myr. \citet{2004ApJ...609..854M} 
discuss in detail several scenarios to explain this discrepancy, and arrive at the 
conclusion that a large age spread is unlikely. Instead, they favour a scenario where
this apparent age offset arises from evolutionary model uncertainties related to 
accretion, deuterium burning and/or convection effects. Thus, except for 
\citet{2003A&A...404..913S}, all recent studies give an age of $\sim 5$\,Myr and no 
convincing evidence for strong age spread. This is the assumption which we will use
in the remainder of this paper. Still there are some concerns, and a future 
study of ages in UpSco based on Lithium absorption is desirable. 

Visual extinctions have been determined for our sample by \citet{2000AJ....120..479A} 
and \citet{2004AJ....127..449M}. They find an upper limit of $A_V\sim 2$\,mag, for most
objects the optical extinction is $\lesssim 1$\,mag. This shows that a) near-infrared
magnitudes are not significantly affected by extinction, and b) there is no indication
for close to edge-on disks from the optical magnitudes. Therefore, we can neglect the
effects of disk inclination for most of the following analysis. 

In Table \ref{targets}, we list effective temperatures for our 36 targets. These values
have been determined by comparing the spectral types with the effective temperature scale 
provided by \citet{2003ApJ...593.1093L}. \citet{2004ApJ...609..854M} derived fundamental
parameters for some of our objects by spectral fitting, and their effective temperatures
tend to be lower than our estimates by as much as 200\,K. This discrepancy reflects the 
difficulties in assigning a reliable effective temperature scale for these cool objects, but 
might also be affected by problems in spectral typing. The absolute uncertainties of the 
effective temperatures in Table \ref{targets} are thus probably in the range of $\pm 200$\,K. 
Since they have been derived in a consistent way, the relative accuracy is certainly better.

The masses for our targets can be constrained by comparing the effective temperatures with 
the 5\,Myr isochrones from \citet{2003A&A...402..701B}. With this approach, we obtain 
masses between $\sim 0.02$ and 0.12$\,M_{\odot}$. The masses derived by
\citet{2004ApJ...609..854M} span a larger range, from 0.01 to 0.26$\,M_{\odot}$, where 
the discrepancies are probably again due to some problems of the evolutionary tracks at 
young ages \cite[see][]{2002A&A...382..563B}. Still, these estimates confirm that all 
our objects have very low masses, although some of them may not be substellar.

\subsection{SEDs: General remarks}
\label{sedgen}

The mid-infrared SEDs for the sample of 35 brown dwarfs in UpSco are plotted in Fig. 
\ref{f9}. In this Section, we give an overview of the derived SEDs purely based on the 
observational results, without invoking models for comparison. 

From the appearance of the SED, we can clearly divide the objects in two classes: There
are thirteen objects with clear excess at 24$\,\mu m$, which also have the highest fluxes 
between 9 and 11$\,\mu m$. The remaining ones are either not detected or have very low fluxes 
in MIPS, and are in most cases faint in IRS. In these two types of SED we likely see objects 
with and without disks, roughly speaking. (We will show in Sect. \ref{diskfreq} more 
definitively that the 13 with the highest 24$\,\mu m$ fluxes are safe disk detections.) 
For all objects with MIPS detections, 
the SEDs (given in flux units) are surprisingly flat: With three exceptions, which will be 
discussed below, the fluxes at 9-11$\,\mu m$ are consistent with the fluxes at $24\,\mu m$, 
and generally on the level of a few mJy. The MIR SED is sensitive to both disk structure 
and inclination \cite[e.g.][]{2002ApJ...567.1183W}; since we can rule out high inclinations 
(i.e. close to edge-on disks) for all our targets based on the low visual extinction (see Sect. 
\ref{objprop}), this points out that most brown dwarf disks in UpSco have similar structural 
characteristics, resulting in flat SEDs.

The three exceptions are usco55, usd161833, and usd161939. These are the only objects with
MIPS and IRS fluxes $>6$\,mJy, and the only ones which show clear excess flux at 24$\,\mu m$, 
compared with $9-11\,\mu m$. Rising flux levels in the mid-infrared are a trade mark of 
strongly flared disks, as shown in numerous comparisons of observed and modeled SEDs \citep[e.g.][]{1987ApJ...323..714K,2004A&A...421.1075D}. When the opening angle of the disk 
increases with its radius, the disk will intercept more light from the central object than a
flat disk, leading to excessive mid-infrared emission. On the other hand, a flat 
disk where the dust has settled to the midplane causes a flat or gradually decreasing 
SED in the mid-infrared. Thus, the three objects with rising SEDs are probably the ones 
with the most flaring in our sample. Over all, the degree of flaring is low in UpSco brown 
dwarf disks, or conversely, most disks are likely affected by dust settling.

All three objects with rising SEDs are known to have companions (see Table \ref{targets}): 
usco55 at $\sim 18$\,AU (0.12") \citep{2005ApJ...633..452K}, usd161833 at $\sim 130$\,AU (0.9") 
\citep{2005ApJ...633L..41L}, and usd161939 at $\sim 25$\,AU (0.08")\footnote{usd161939 has 
additionally a wide companion which is, however, likely a background source \citep{2006A&A...451..177B}.} 
\citep{2006A&A...451..177B}. (All physical separations given in AU are projected values assuming a 
distance of 145\,pc; because of the projection factor strictly speaken they should be considered 
as lower limits to the true separation.) The two closer companions are far from being 
resolved with MIPS and IRS (which have pixel sizes of 2.4 and 1.8" respectively). The 240\,AU 
companion of usd1618333 is not resolved by MIPS either, and it is likely to contaminate the IRS 
spectrum to some extent. Thus, our MIPS fluxes for these three objects include the flux from the 
companion. Photospheric fluxes from the companions, however, are not able to explain the 'flared' 
SED appearance, since the photospheres cannot contribute more than 0.5\,mJy, less than 5\% of our 
total 24$\,\mu m$ fluxes for these three objects. Moreover, companions also contribute to near-infrared 
and IRS fluxes, thus cannot produce an increasing SED. Therefore, we attribute the excess emission 
at 24$\,\mu m$ to their disks. A discussion of possible influences of binarity on the disks follows 
in Sect. \ref{bin}.

Comparable to the overall SED, the IRS spectra alone appear flat or decreasing
in most cases. Only three objects -- usco112, usd160958, usd161939 -- show an indication of
an emission feature (further investigated in Sect. \ref{gwen} and Fig. \ref{f4}). Emission 
features around 10$\,\mu m$ (often referred to as silicate features) are assumed to have its 
origin in an optically thin hot surface layer of small dust grains. The absence or weakness 
of this feature thus can be interpreted as lack of hot, small grains -- possibly due to grain 
growth and subsequent dust settling to the (colder) disk midplane. Dust settling is confirmed 
for the majority of our objects by the general appearance of the SED (see above). Additional 
factors which can lead to dust settling and thus may weaken the silicate feature are a low 
gas-to-dust ratio or weak turbulence in the disk.  We caution that disk inclination can also 
weaken the emission feature and thus might also play a role in some of our targets, although 
high inclinations are unlikely (see Sect. \ref{objprop}). For a more detailed discussion of
the minimal effects of inclination for our sample see also Fig. \ref{f8} and the assciated
discussion in Sect. \ref{kenny}.

Both findings, the uniformly flat SEDs and the paucity of strong dust emission features,
demonstrate that UpSco brown dwarfs harbour disks that are clearly in an advanced stage of
their evolution. At an age of 5\,Myr, flaring is not a major trait anymore. Instead, dust 
settling likely affects the majority of the disks, possible accompanied (and caused) by 
grain growth. We will come back to these findings in Sect. \ref{disc}, where we carry out
comparisons with disk properties for younger and more massive objects.

\subsection{Disk frequency}
\label{diskfreq}

\begin{figure}
\includegraphics[angle=-90,width=13.8cm]{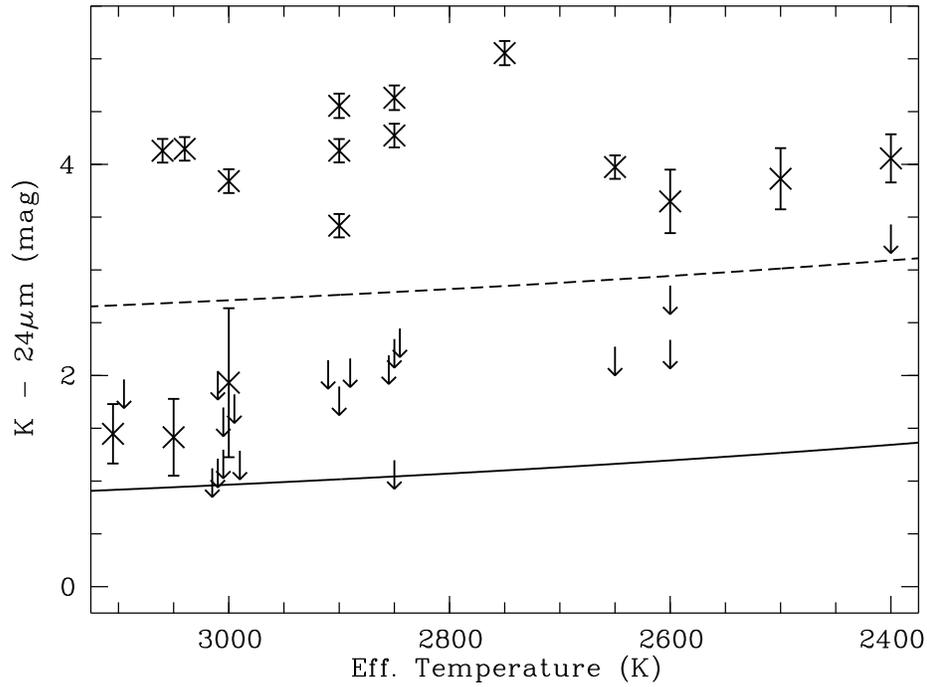}
\caption{K-band minus 24$\mu m$ colours for UpSco brown dwarfs calculated from MIPS fluxes vs.
effective temperature. Upper limits correspond to 2$\sigma$ upper limits at 24$\mu m$. The
solid line show the photospheric colours estimated by assuming blackbody radiation; the
dashed lines show colours assuming five times the photospheric emission at 24$\mu m$ and 
thus significant excess (see Sect. \ref{diskfreq}).  \label{f1}}
\end{figure}

In Fig. \ref{f1} we plot the $K - 24\,\mu m$ colour vs. effective temperature for all
targets. This plot will be used to derive an estimate for the disk frequency in our sample. 
Specifically, we want to find out for how many objects the measured MIPS flux (or upper limit) 
is consistent with photospheric emission. For this purpose, we calculate blackbody fluxes for the 
given effective temperature range at 2.2 and 24$\,\mu m$, from which we obtain the photospheric 
$K - 24\,\mu m$ colour.

The solid line in Fig. \ref{f1} shows the result: Photospheric colours for
our targets are between 0.9 and 1.4\,mag, depending on effective temperature.
Dashed lines show the colour assuming five times the photospheric emission 
at 24\,$\mu m$. All objects with colours exceeding this level, have strong excess emission,
and are thus likely to harbour a disk. Thirteen objects with MIPS detection are in this
colour range, and are thus primary disk candidates. For one more object, the upper 
limit is too high to exclude excess due to a disk. For the remaining objects, the MIPS 
fluxes are fully consistent with pure photospheric emission. In total, the disk frequency 
for our sample is $37\pm 9$\% (13 out of 35, uncertainty corresponds to the 1$\sigma$
confidence interval based on binomial statistics), but might be as high as 40\% (14 out of 
35), if we count the upper limit above the threshold.

This estimate is quite robust and does not depend on distance and radii of the objects.
Using a conservative limit (dashed line), we make sure that uncertainties in effective
temperatures, which might be considerable (see above), and the inaccuracy of the black
body approximation does not strongly affect our result. This conservative approach also 
prevents us from confusing disks with non-resolved, cool companions, which might enhance 
the mid-infrared flux. 

As can be seen in Fig. \ref{f1} there is a clear dichotomy 
in the mid-infrared colours of our targets: Down to $T_{\mathrm{eff}} = 2600$\,K, objects with 
strong $24\mu m$ excess are separated by at least 1.2\,mag (i.e. a factor of three) from objects
without excess. (This is a lower limit, because most objects without excess are in fact
non-detections.) For the lowest mass objects, where the noise level increases, the difference 
between both populations is still at least a factor of two in colour. Thus, most of the objects 
with disks are clearly separated from objects without disks, leaving only a small number of 
faint objects with ambiguities. Thus, we can quite reliably conclude that, based on the $24\,\mu m$ 
photometry, the disk frequency in our sample is 37\%. This value, derived from an unbiased survey,
is somewhat smaller, but still consistent with the disk frequencies obtained from smaller 
samples of brown dwarfs in UpSco \citep[50\%, see][]{2003AJ....126.1515J,2006astro.ph..8395B}.

\begin{figure}
\includegraphics[angle=0,width=5.4cm]{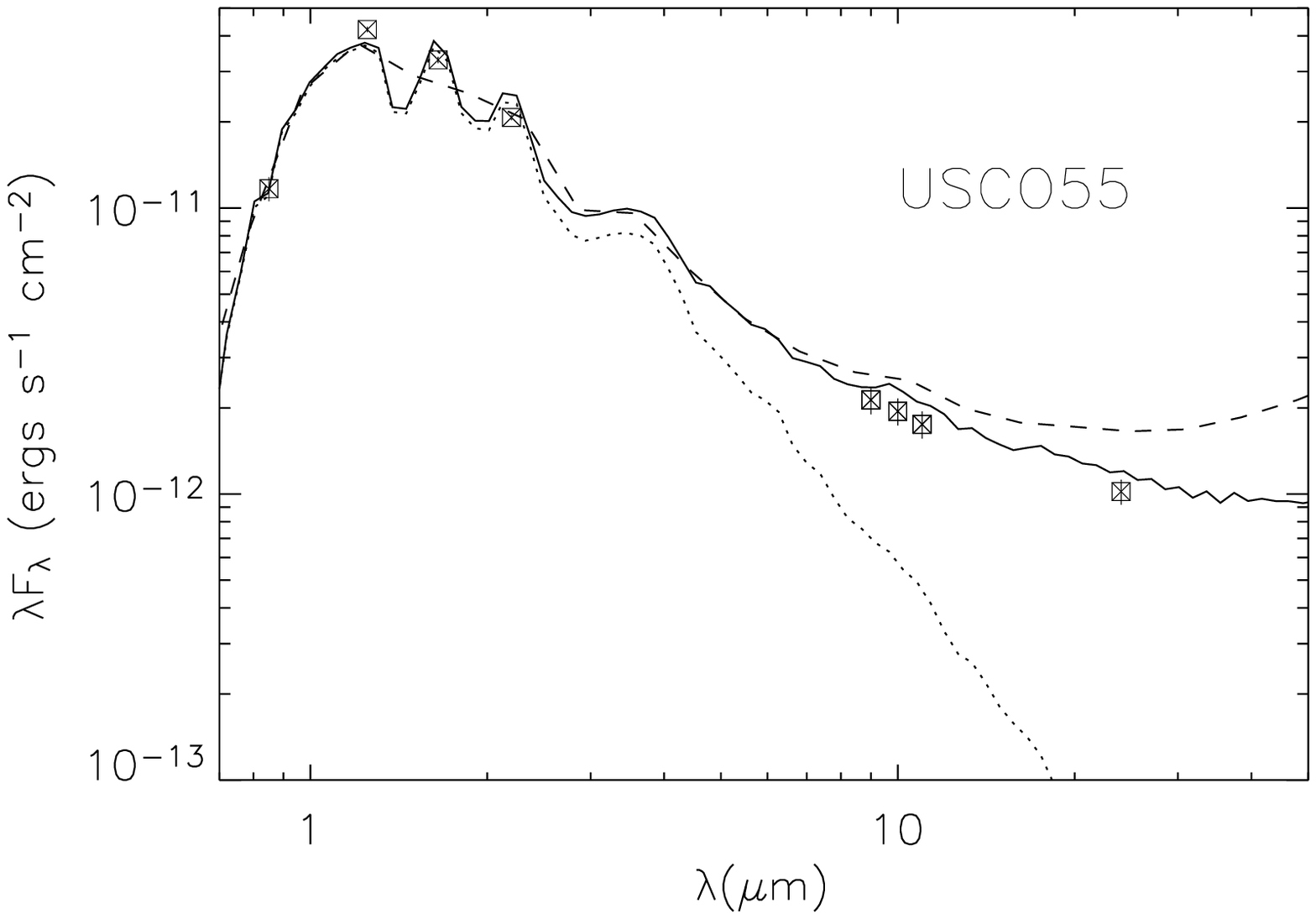}
\includegraphics[angle=0,width=5.4cm]{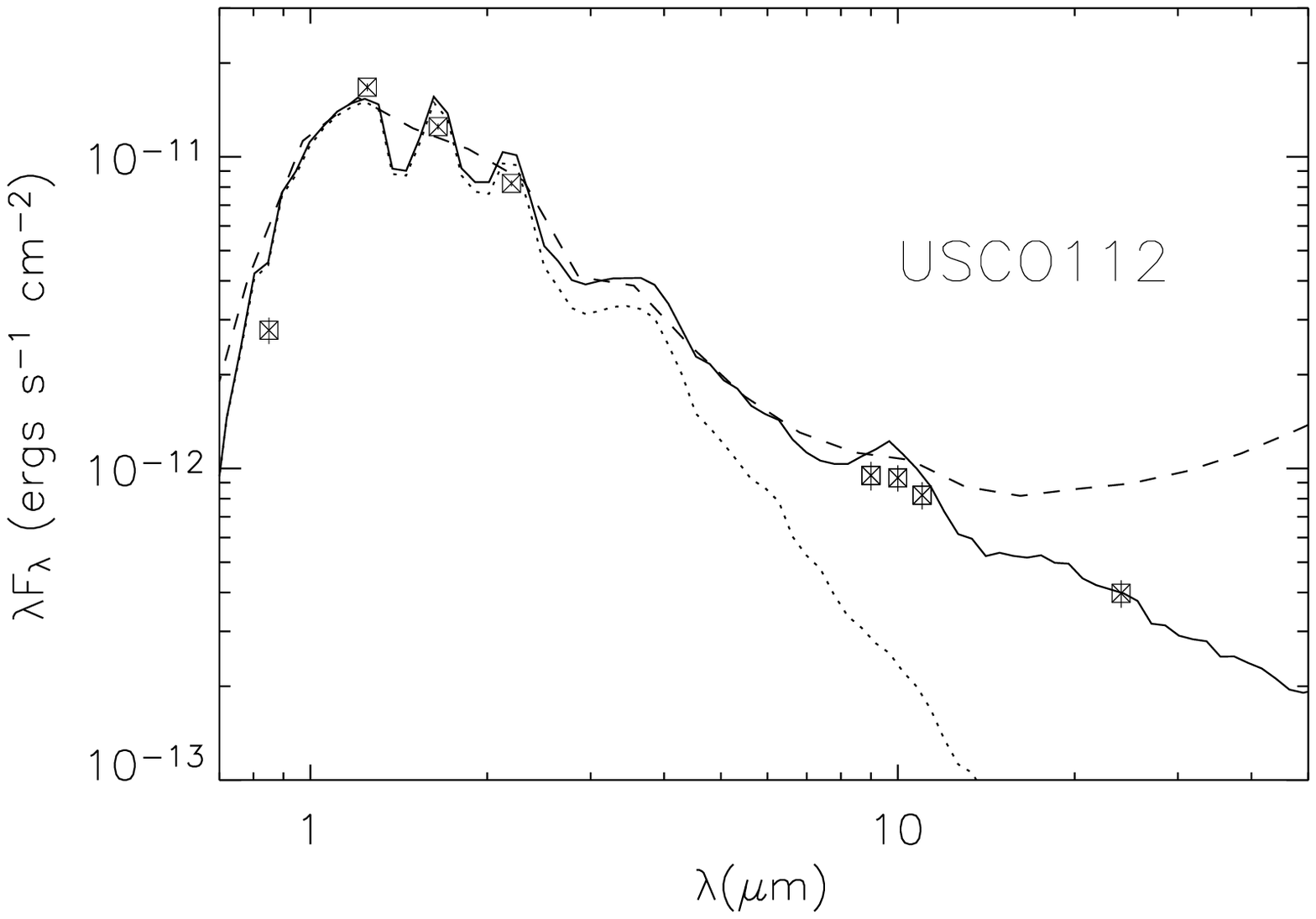}
\includegraphics[angle=0,width=5.4cm]{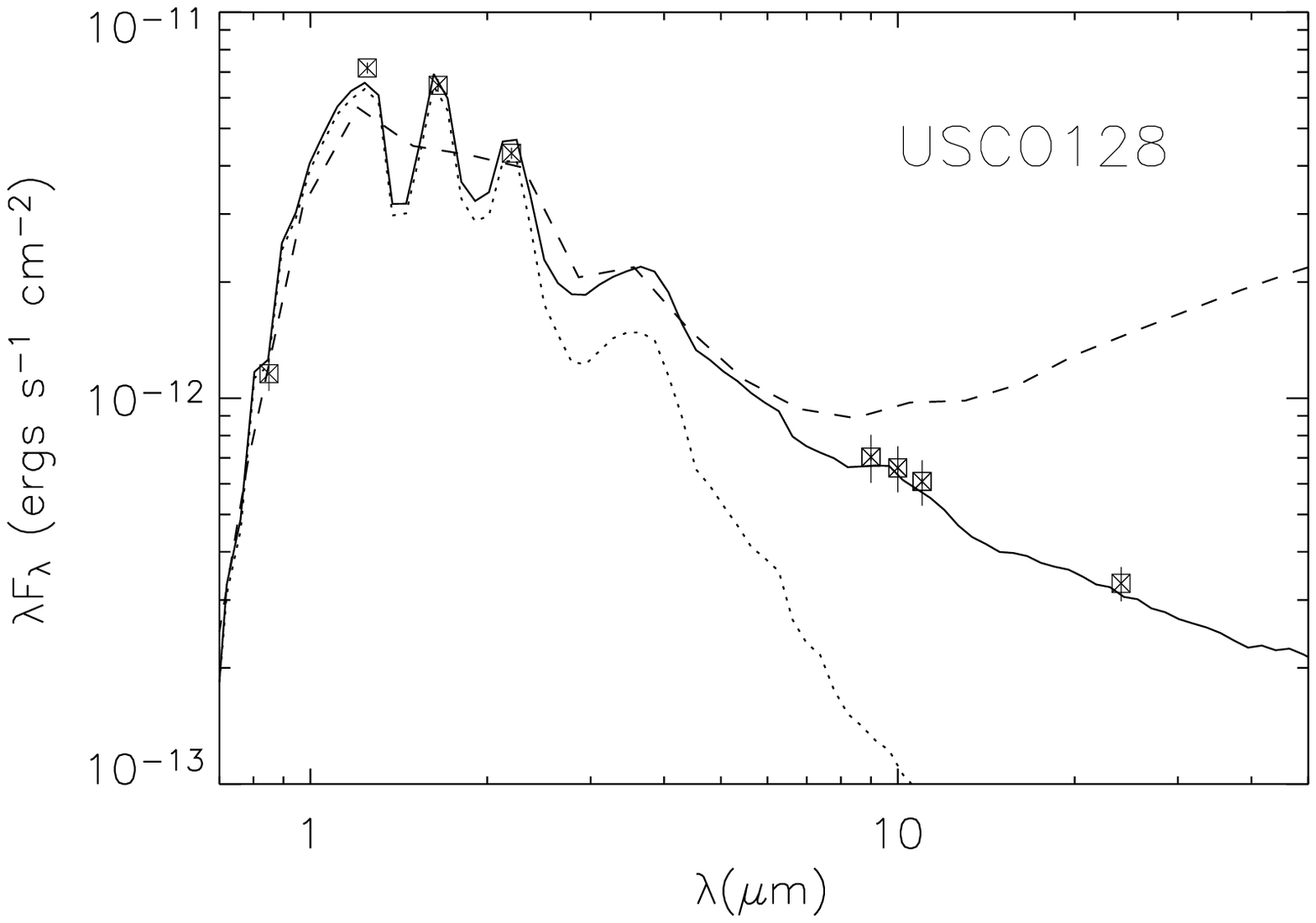}\\
\includegraphics[angle=0,width=5.4cm]{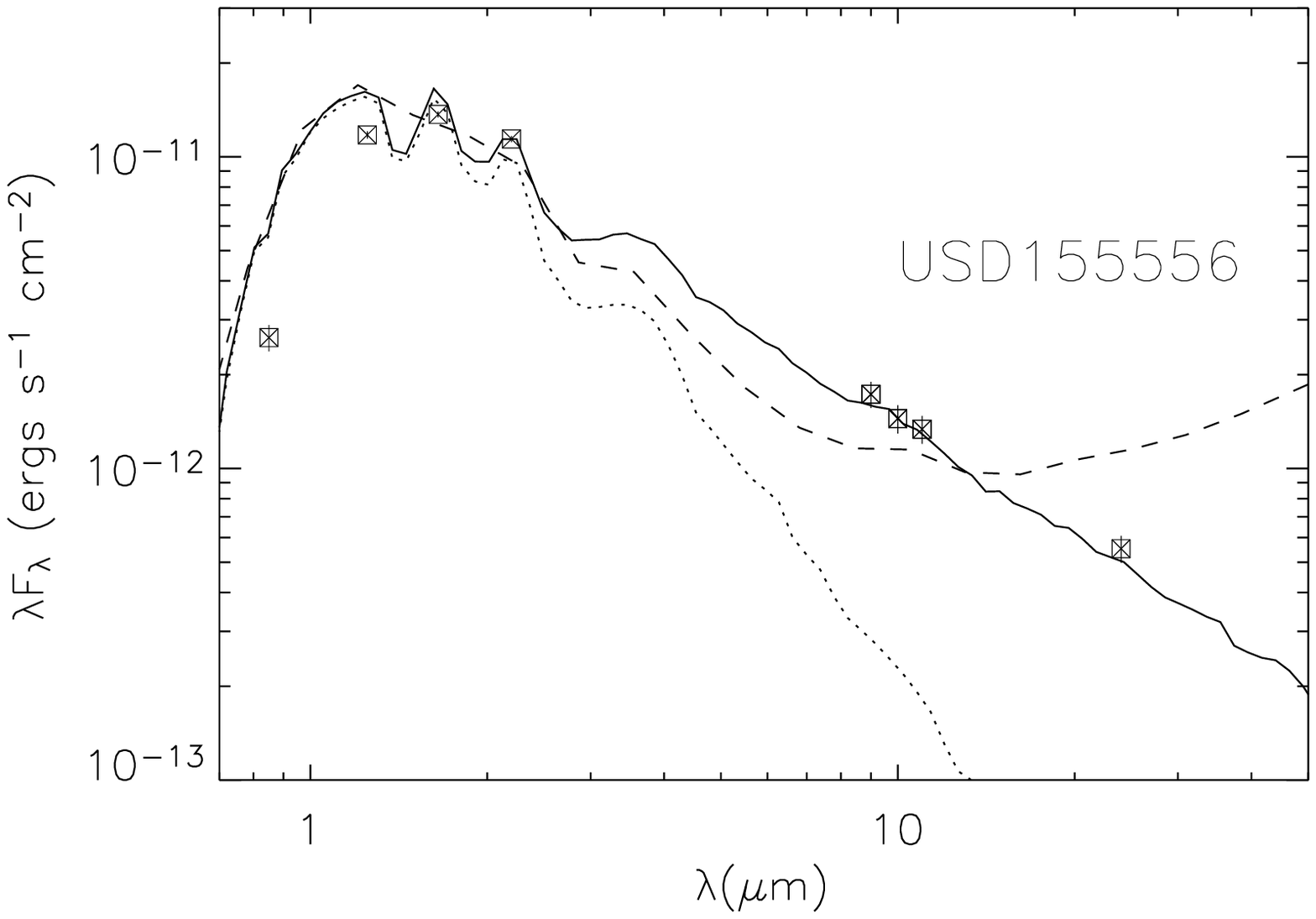}
\includegraphics[angle=0,width=5.4cm]{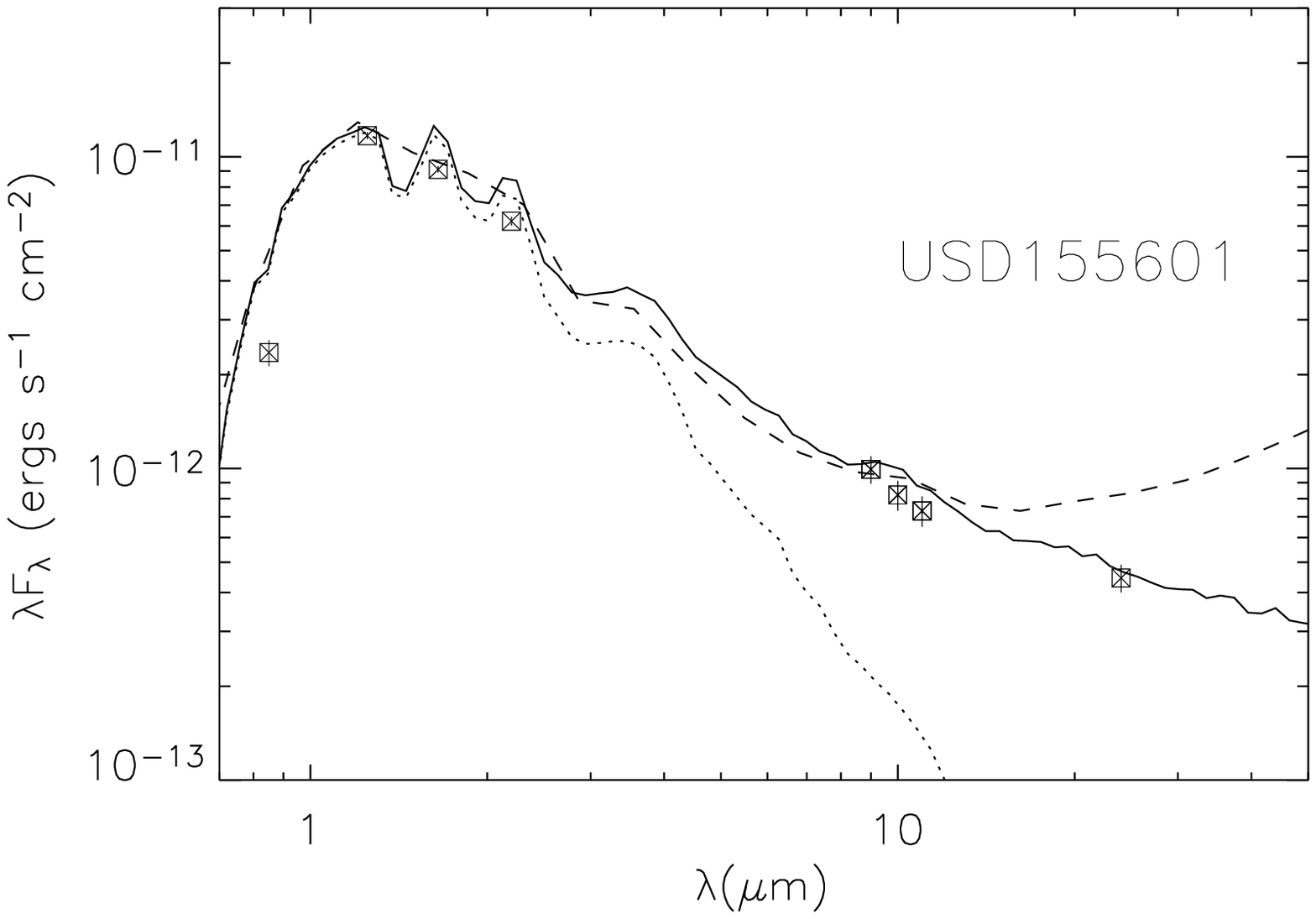}
\includegraphics[angle=0,width=5.4cm]{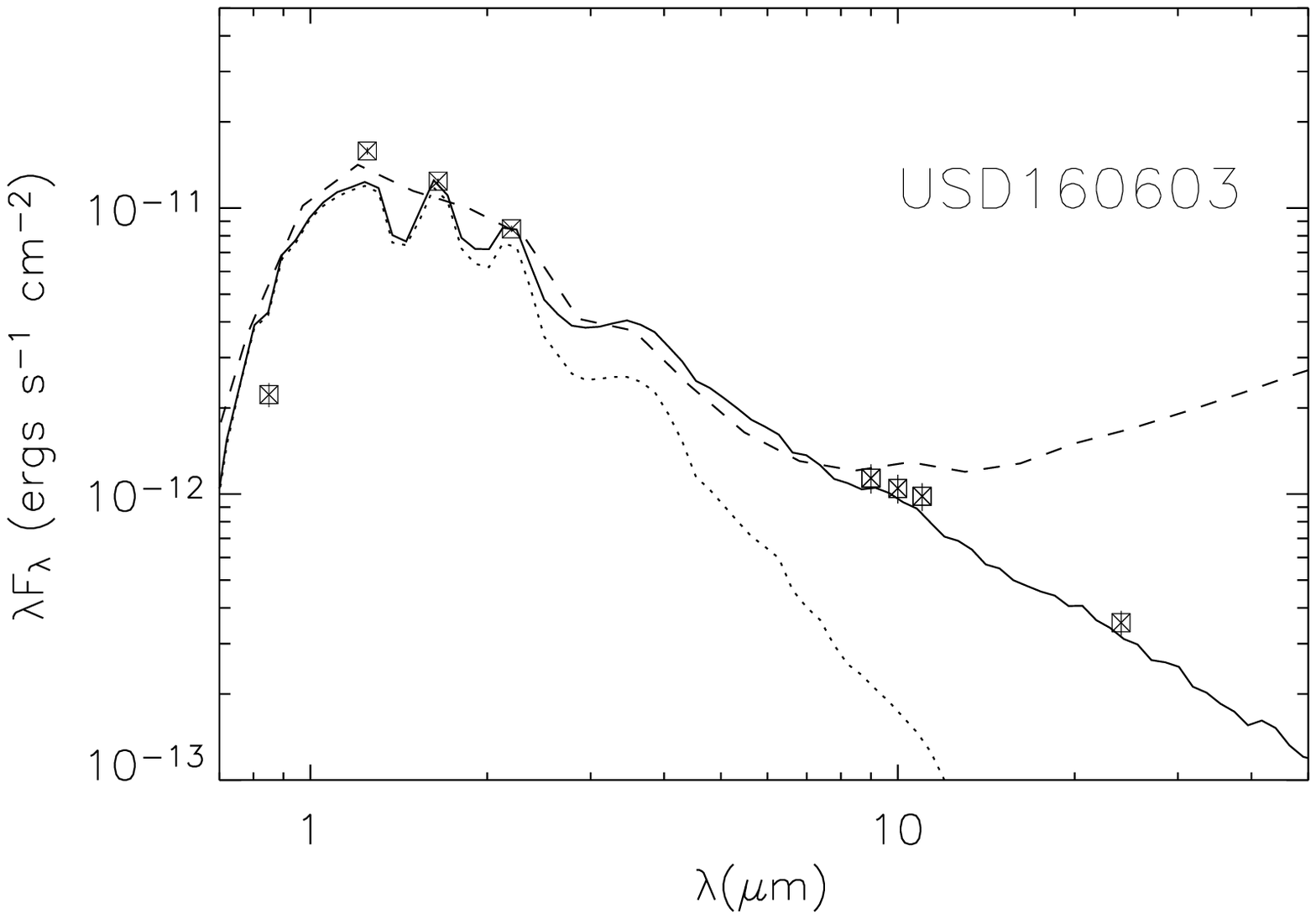}\\
\includegraphics[angle=0,width=5.4cm]{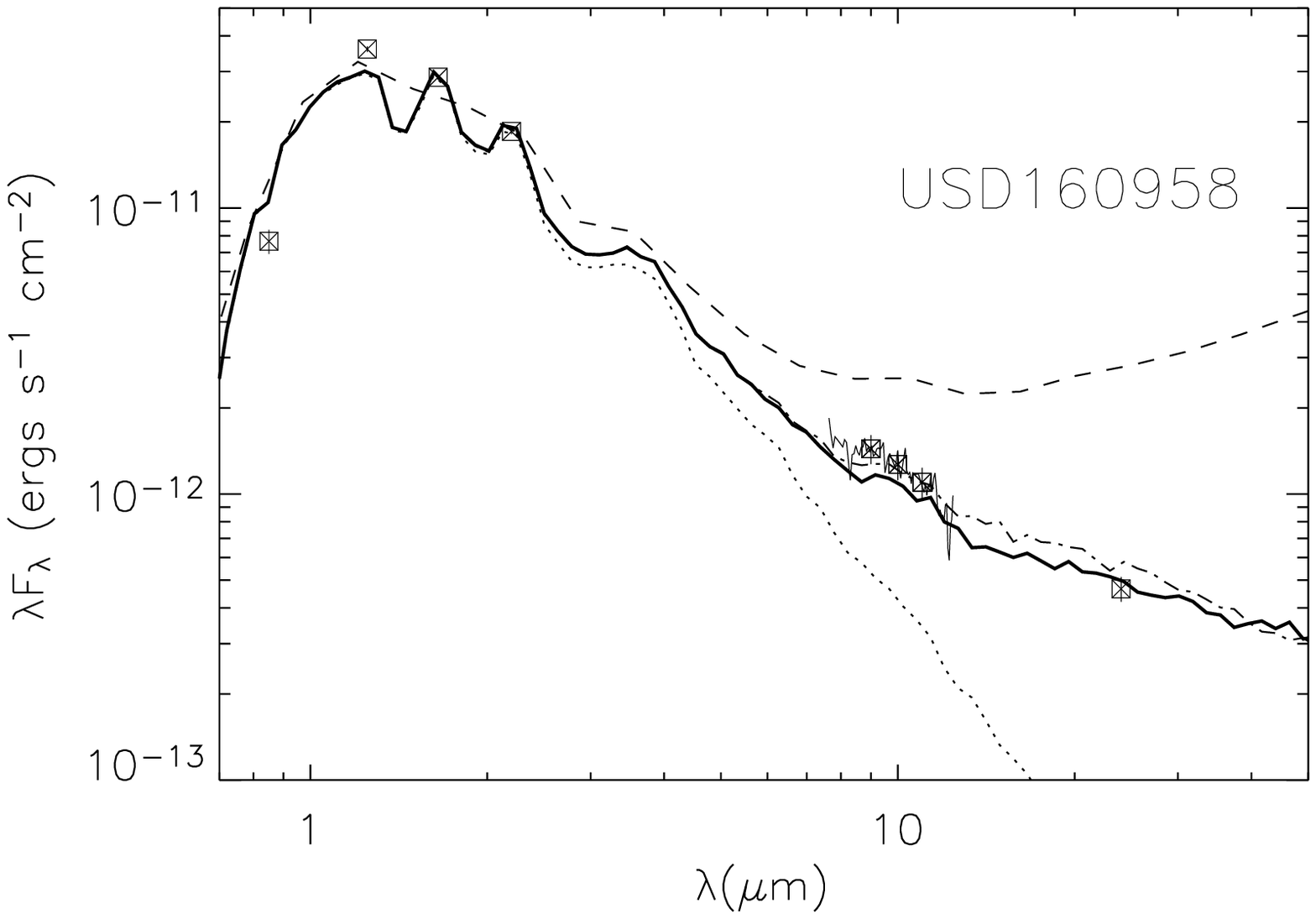}
\includegraphics[angle=0,width=5.4cm]{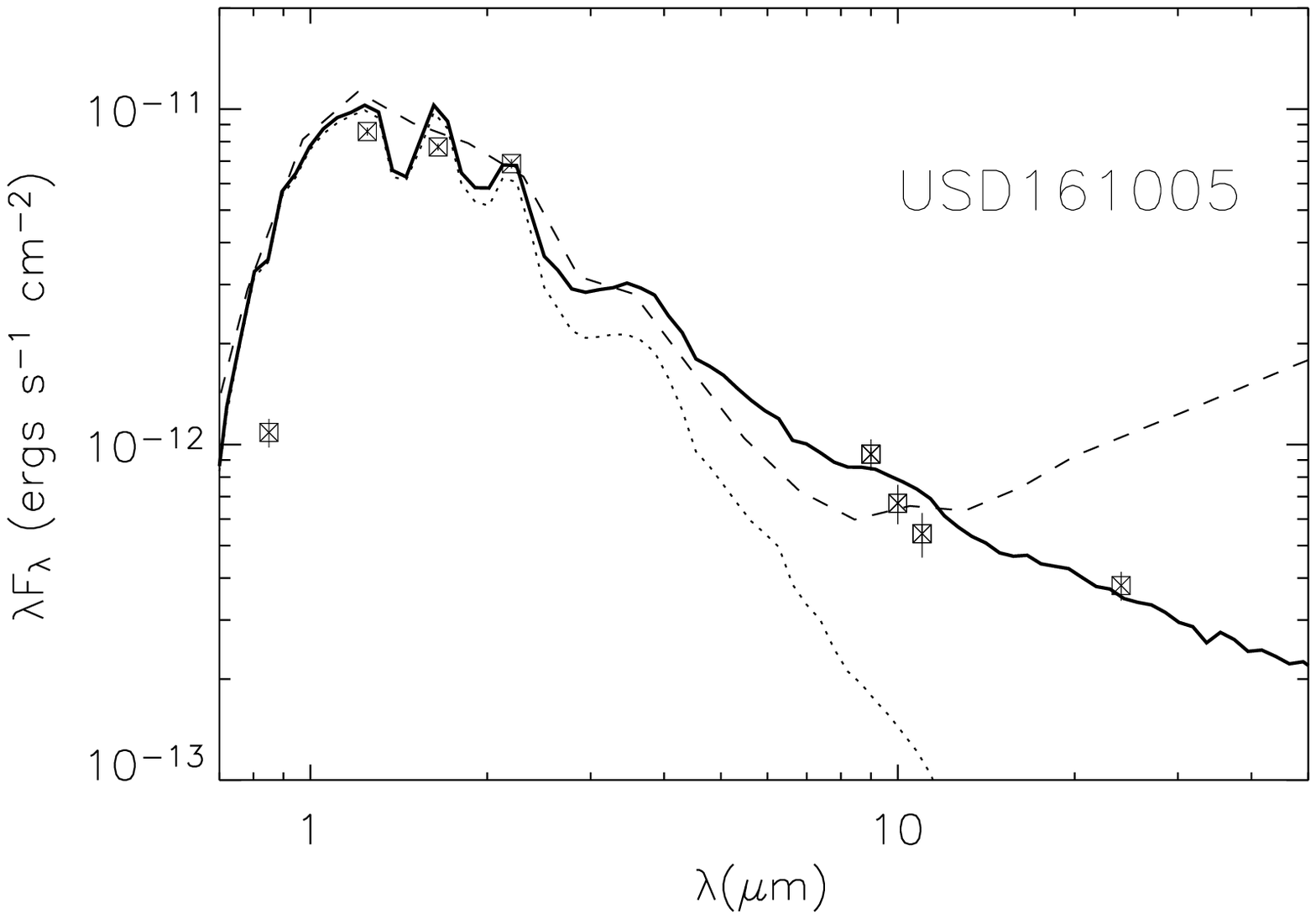}
\includegraphics[angle=0,width=5.4cm]{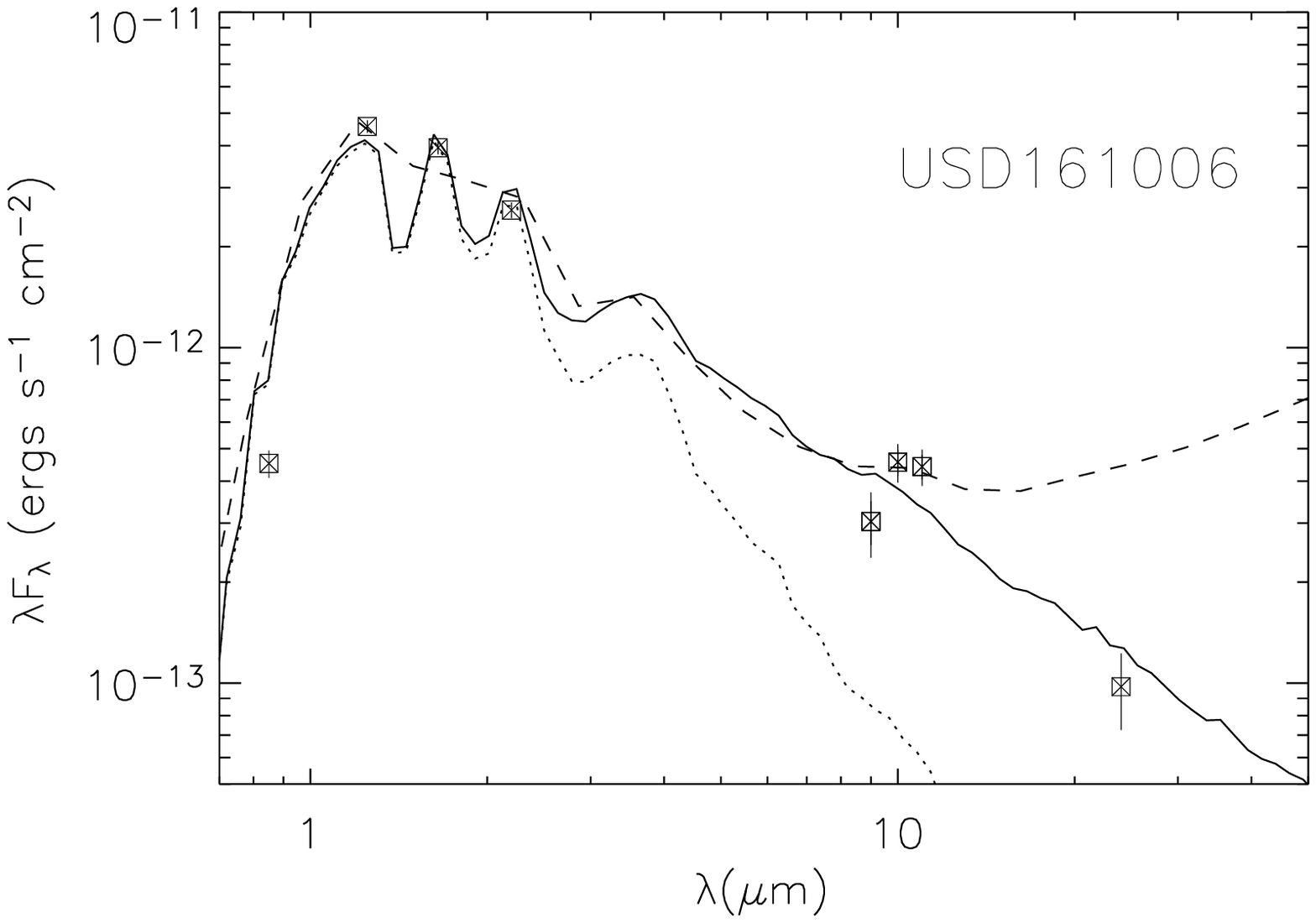}\\
\includegraphics[angle=0,width=5.4cm]{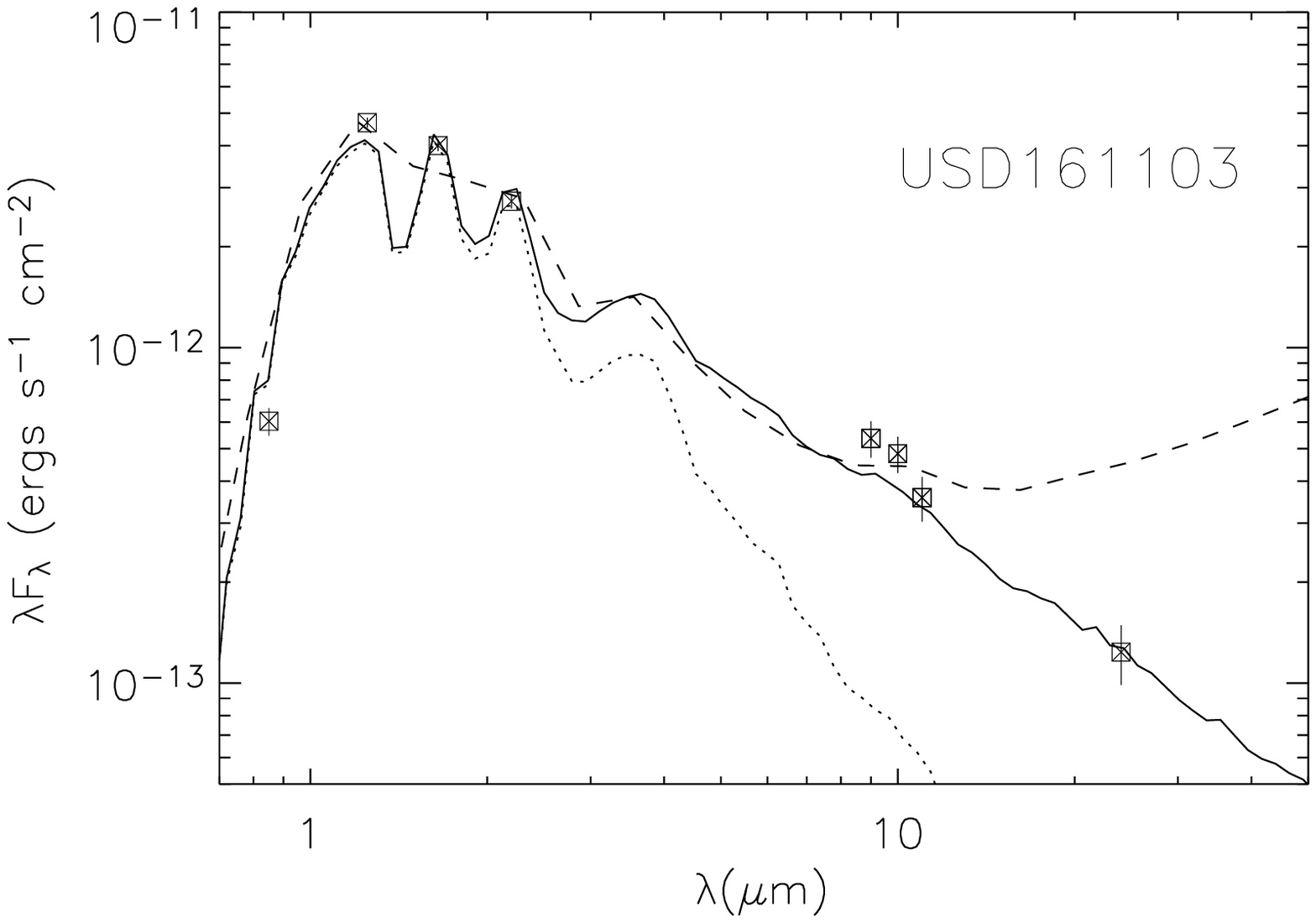}
\includegraphics[angle=0,width=5.4cm]{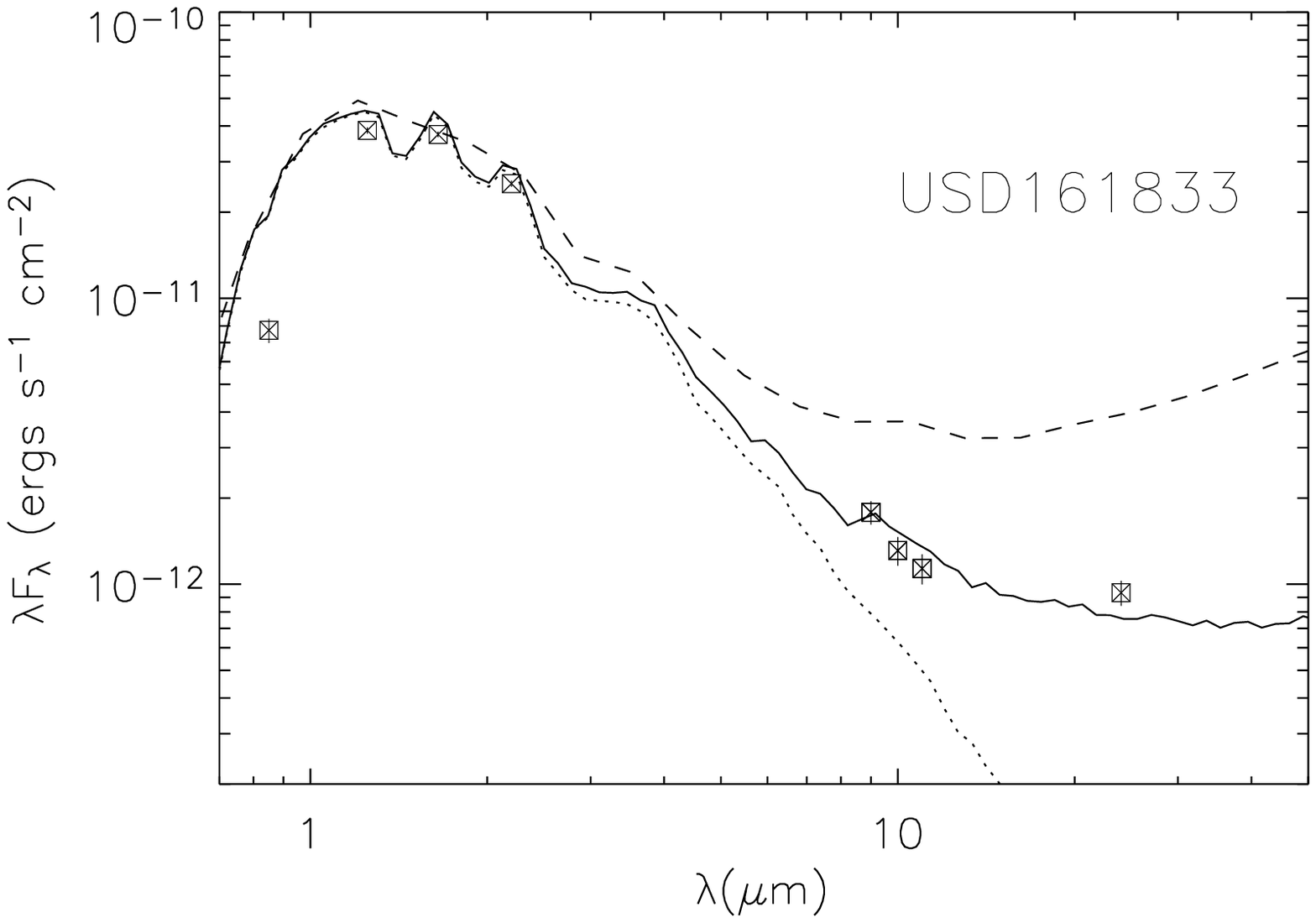}
\includegraphics[angle=0,width=5.4cm]{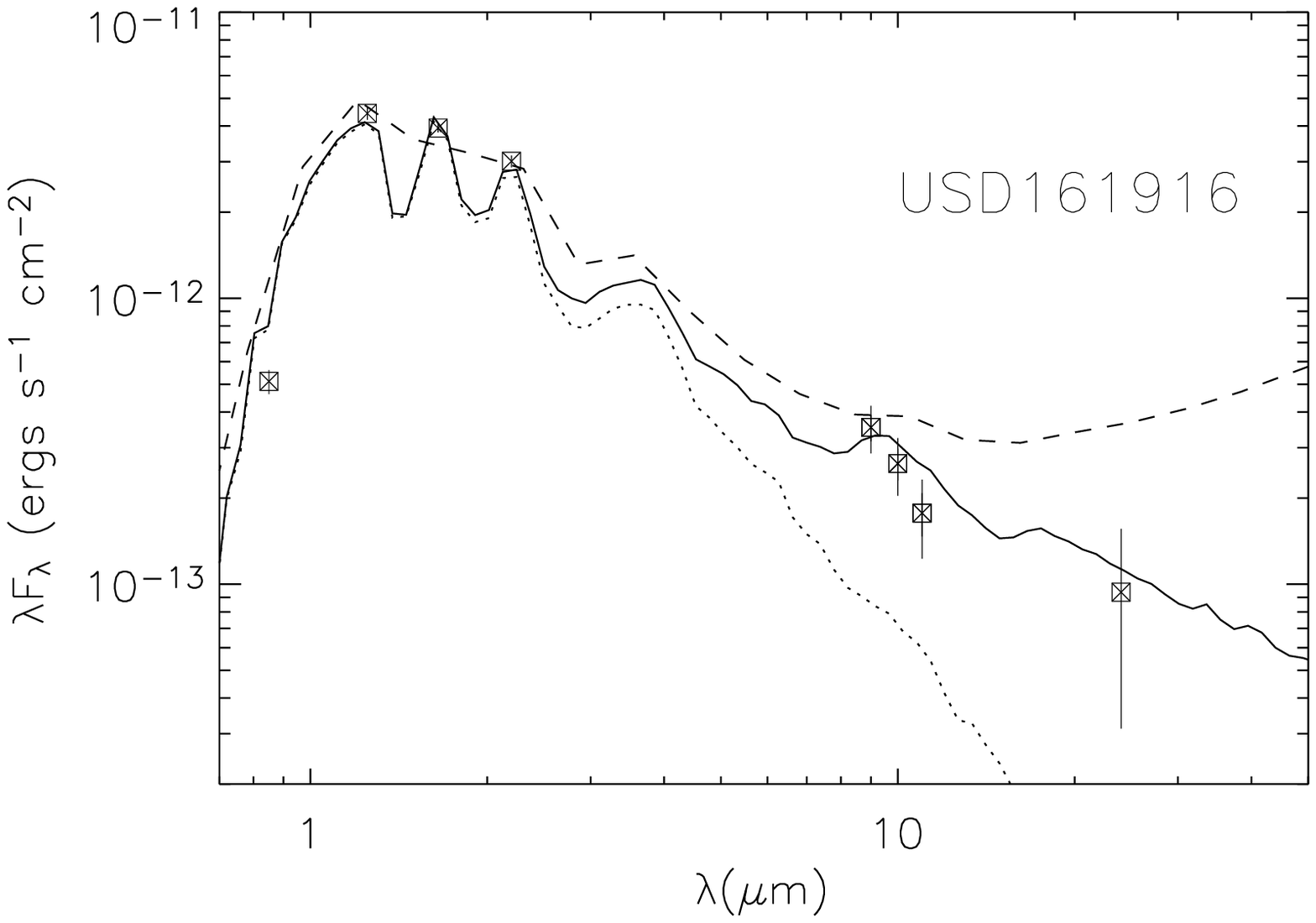}\\
\includegraphics[angle=0,width=5.4cm]{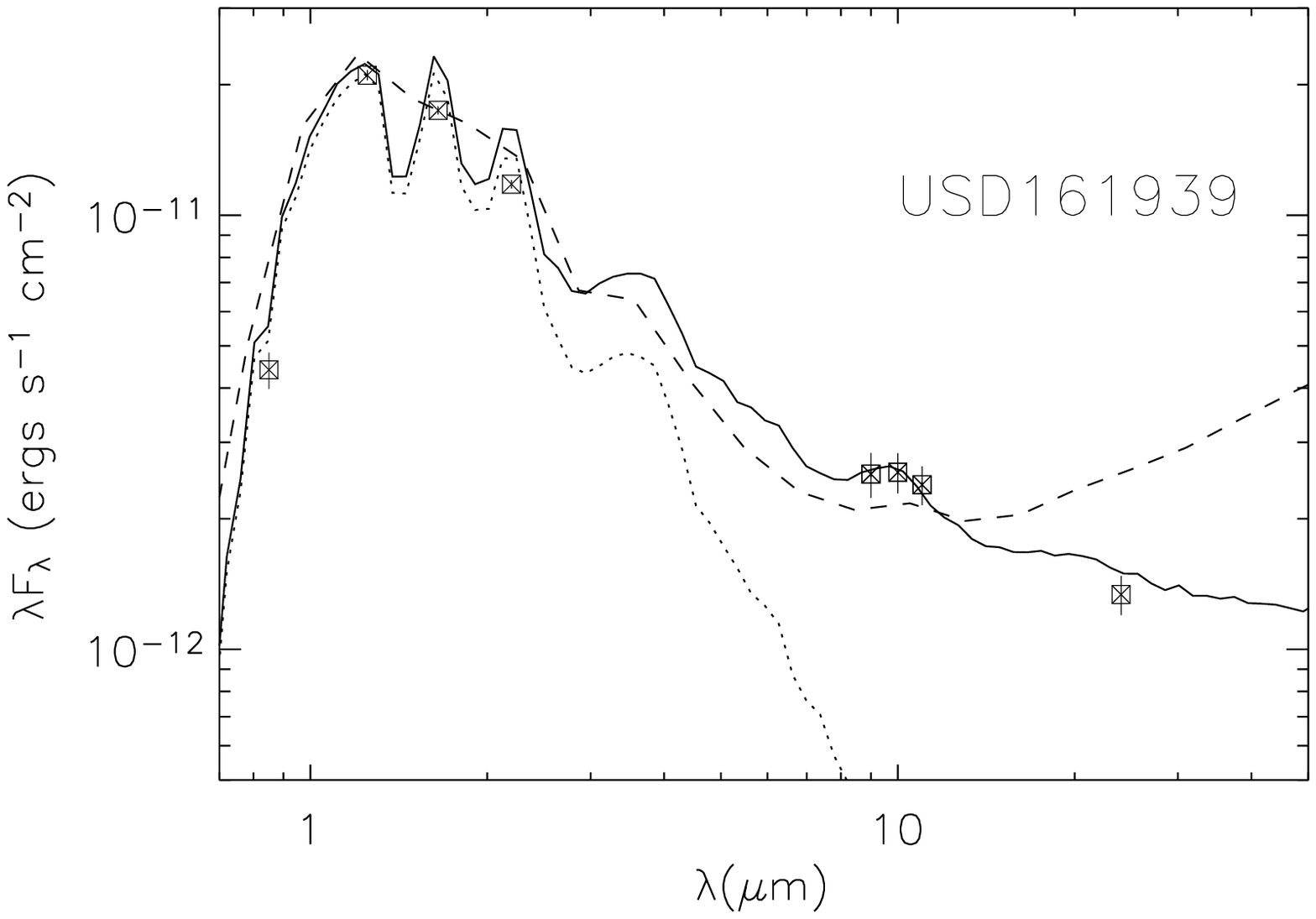}
\caption{Spectral energy distribution for objects with MIR excess emission in comparison 
with models (dotted lines: photospheric flux, dashed lines: photosphere + hydrostatic disk, 
solid lines: photosphere + two-component disk, see Sect. \ref{kenny}). For usd160958, a model
with disk radius of 5\,AU is shown as dash-dotted line (7th panel). The parameters for
the models shown in solid lines are listed in Table \ref{modelparams}. \label{f2}}
\end{figure}

All 13 objects with MIPS detection and 
clear excess emission, i.e. objects with safe disk detection, are marked in the last column in 
Table \ref{targets}. For clarity, they are also plotted with different symbols in Fig. \ref{f9}.
Those are the primary candidates for the comparison with models, which will be described in Sect. 
\ref{models}. Their SEDs are shown in Fig. \ref{f2}.

\section{Modeling the SEDs}
\label{models}

In this section, we will compare the observed SEDs with models to obtain more quantitative 
constraints. Two different and independent approaches are used: a) a Monte Carlo radiative transfer
model providing predictions for the MIR fluxes (Sect. \ref{kenny}, and b) linear combinations of 
mass absorption coefficients to fit the dust emission features around 10$\,\mu$m (Sect. \ref{gwen}). 
Both approaches give results consistent with the more qualitative discussion in Sect. \ref{sedgen}: 
Brown dwarf disks in UpSco are ubiquitously affected by dust settling, possibly grain growth, and
inner disk clearing.

\subsection{The overall SED}
\label{kenny}

We adopt the same modeling approach for the UpSco Spitzer data as we did in our 
previous paper on brown dwarf disks \citep{2006ApJ...645.1498S}. These SED models use 
our Monte Carlo radiation transfer code, ignore accretion, use NextGen model 
atmospheres for the input spectra \citep{2001ApJ...556..357A,1999ApJ...512..377H}, 
and assume the disk surface density profile is given by a power law, 
$\Sigma(R)\sim R^{-1}$. The disk density is parameterized by 

\begin{equation}
\rho=\rho_0 \left ({R_\star\over{\varpi}}\right )^{\alpha}
\exp{ -{1\over 2} [z/h( \varpi )]^2  }
\; ,
\end{equation}

where $\varpi$ is the radial coordinate in the disk mid-plane and the scaleheight increases with 
radius, $h=h_0\left ( {\varpi /{R_\star}} \right )^\beta$. We vary the degree of flaring within 
the geometric disk models by adjusting the values of $\beta$ and $h_0$. We assume two dust size 
distributions are present within the disk, a) small grains as with an interstellar-like size 
distribution \citep[][typical grain size $\sim 0.01-0.1\,\mu m$]{1994ApJ...422..164K} and b) larger
grains with sizes up to 1\,mm following a distribution we have used previously to model SEDs of disks 
around classical T Tauri stars \citep[see][for more details]{2002ApJ...567.1183W}. The smaller, 
ISM-like grains have a larger scaleheight than a population of larger grains. This is a simple 
way to mimic dust coagulation and the settling of larger particles towards the disk midplane. 
In our parameterization we assign the mass $M_d$ to the large grains and 
the mass $f_{\rm ISM} M_d$ to the ISM-like grains. The dust mass is then $(1+f_{\rm ISM})M_d$. For 
the conversion to the total disk mass, we assumed a gas to dust ratio of 100. For all models we assume 
that dust in regions close to the star is destroyed if temperatures rise above 1600\,K 
\citep{1996A&A...312..624D}. This condition provides a minimum inner dust radius of typically 
$\sim 6\,R_\star$. Any remaining dust within this gap we assume to be optically thin and therefore 
we effectively have an opacity gap in the disk \citep{1992ApJ...393..278L}. For further details 
of the model ingredients, assumptions, and dust properties see \citet{2006ApJ...645.1498S}.

In our previous paper we discussed the many degeneracies in disk parameters that are inherent 
in SED modeling. In this paper we set the disk mass to be $4.5\times 10^{-4}M_\odot$.  
This disk mass is typical for the sources with 1.3\,mm data that we modeled in our 
Taurus-Auriga sample. In the absence of mm data we cannot constrain the disk mass, as the 
mid-IR SED is not very sensitive to disk mass \citep[e.g.][]{2002ApJ...567.1183W}. The lack of
long wavelength data (sub-mm and mm) also precludes us from determining the optical depths 
of the disks in our sample. As shown by \citet{2004MNRAS.351..607W}, low mass disks in 
hydrostatic equilibrium can provide acceptable fits for the mid-IR SEDs of brown dwarf disks. 
However, our models for Taurus brown dwarf disks showed that such low mass disks are not 
consistent with mm data. Therefore we believe that very low mass, optically thin disks are 
unlikely, but mm data is required to test this further. As we demonstrated in 
\citet{2006ApJ...645.1498S} our data is not sensitive to the outer disk radius, so we choose a 
disk radius of $R_d = 100$~AU. For the object usd160958, which has a close companion at 
$\sim 12$\,AU and a disk (see Table \ref{targets}), we also tried a smaller disk radius of 5\,AU. 

\begin{figure}
\includegraphics[angle=0,width=8cm]{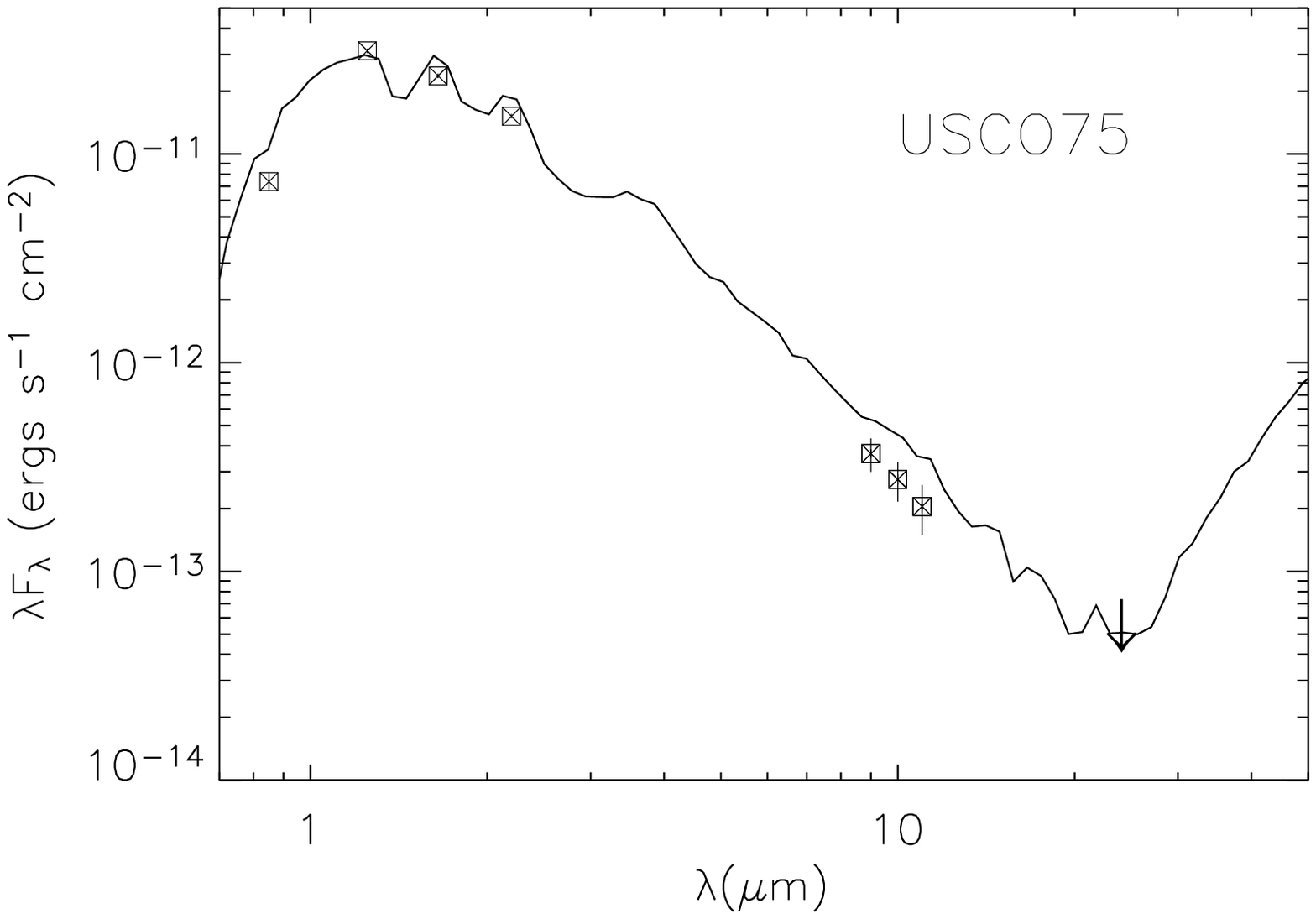}
\includegraphics[angle=0,width=8cm]{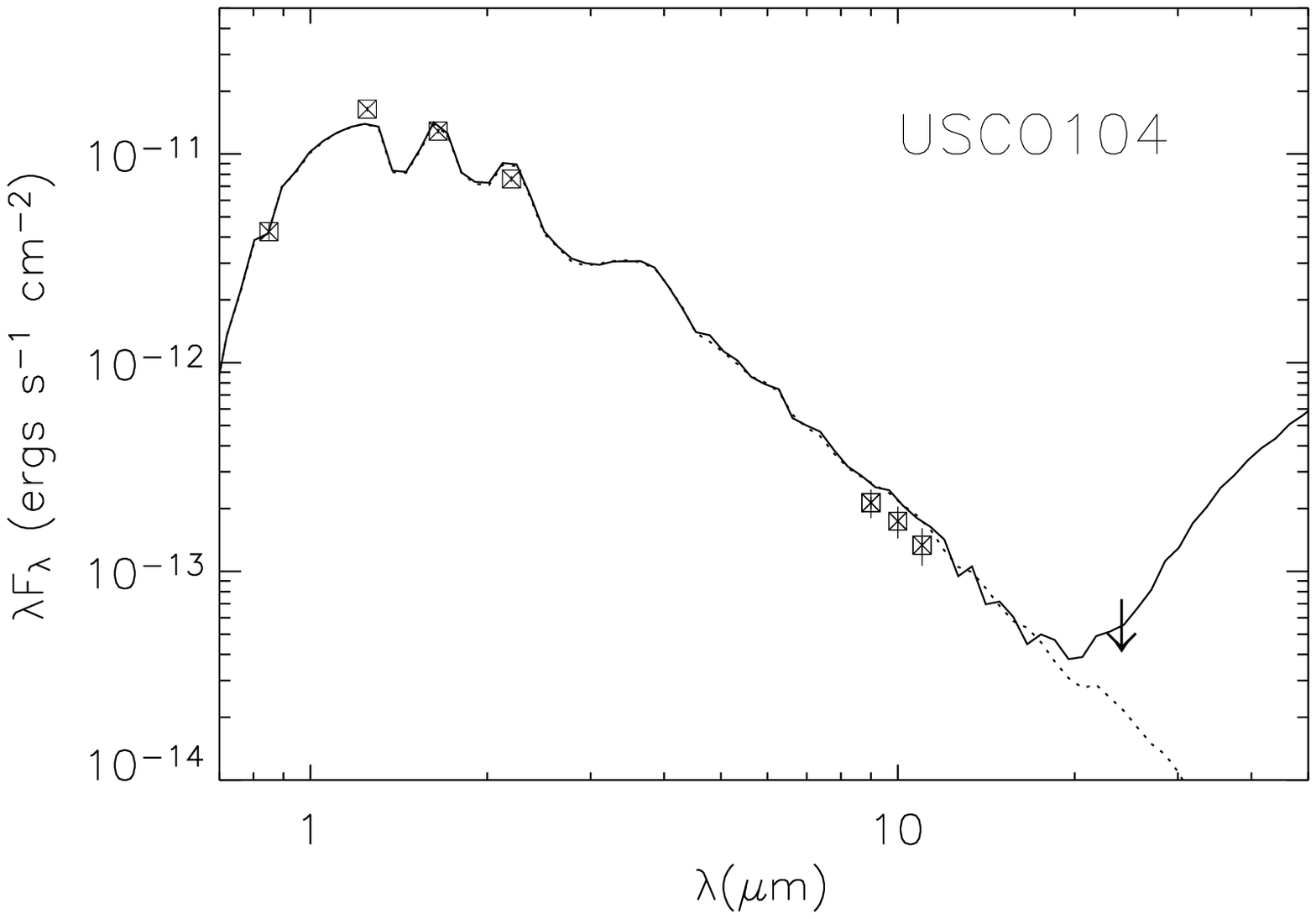}\\
\includegraphics[angle=0,width=8cm]{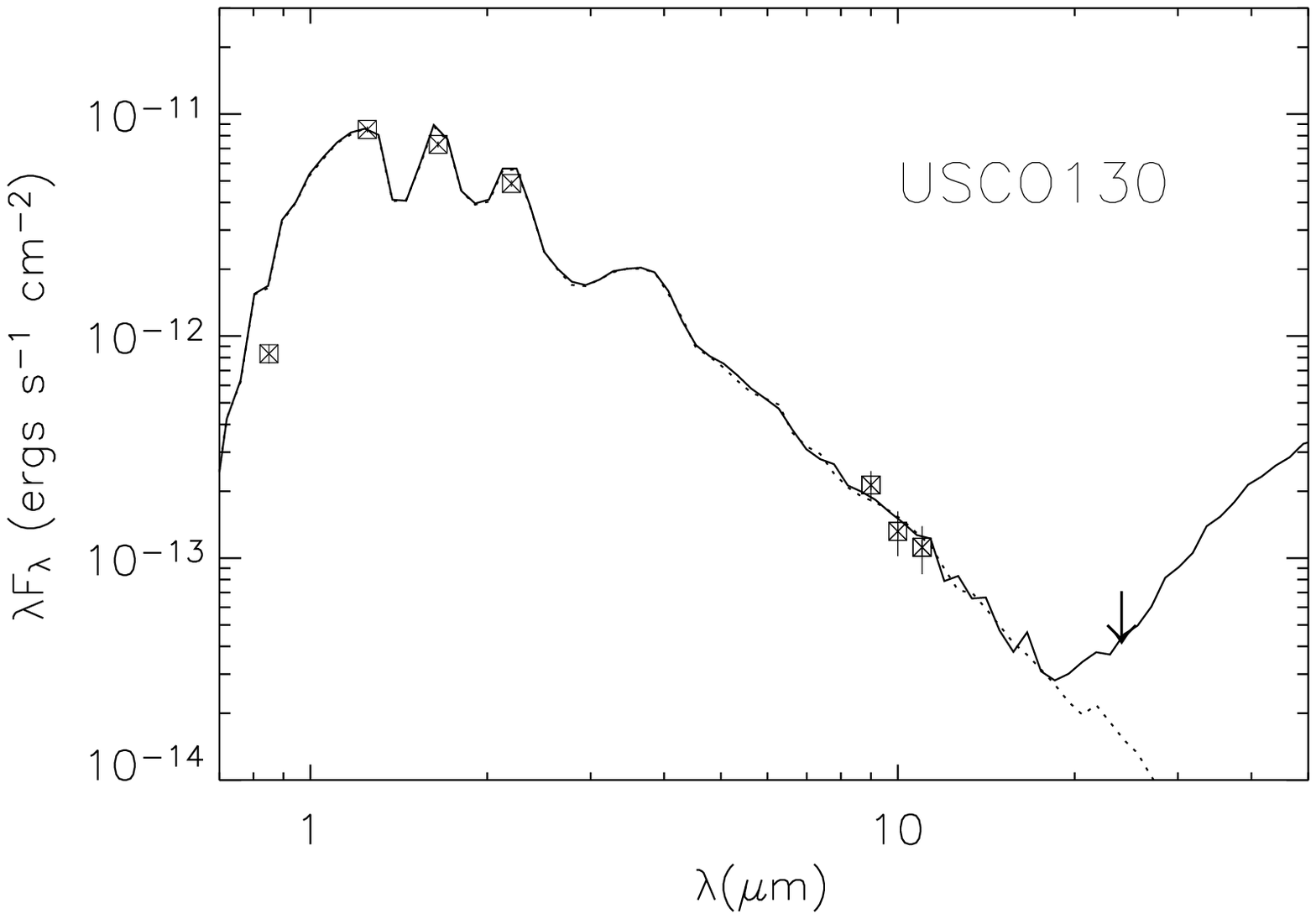}
\includegraphics[angle=0,width=8cm]{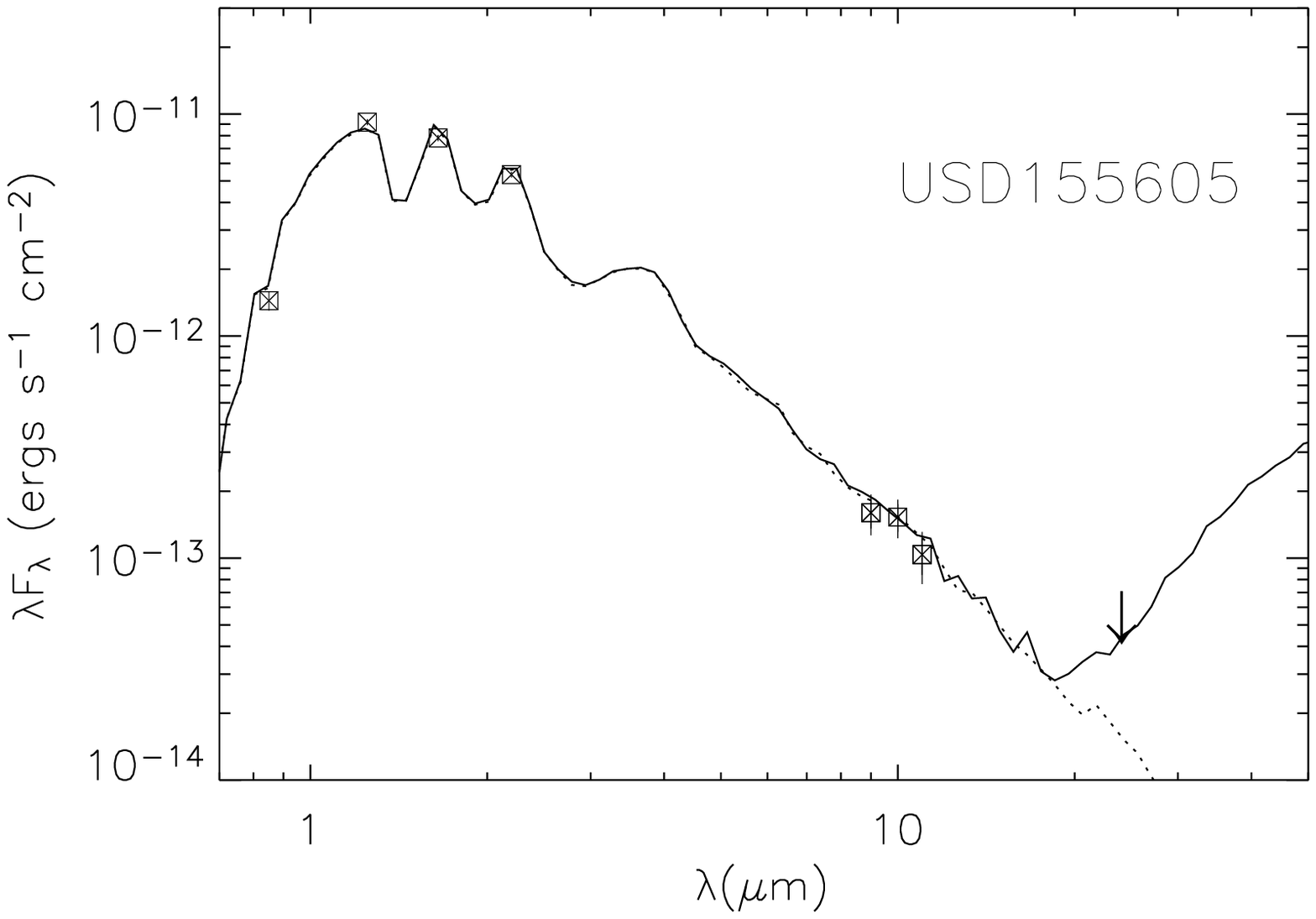}\\
\includegraphics[angle=0,width=8cm]{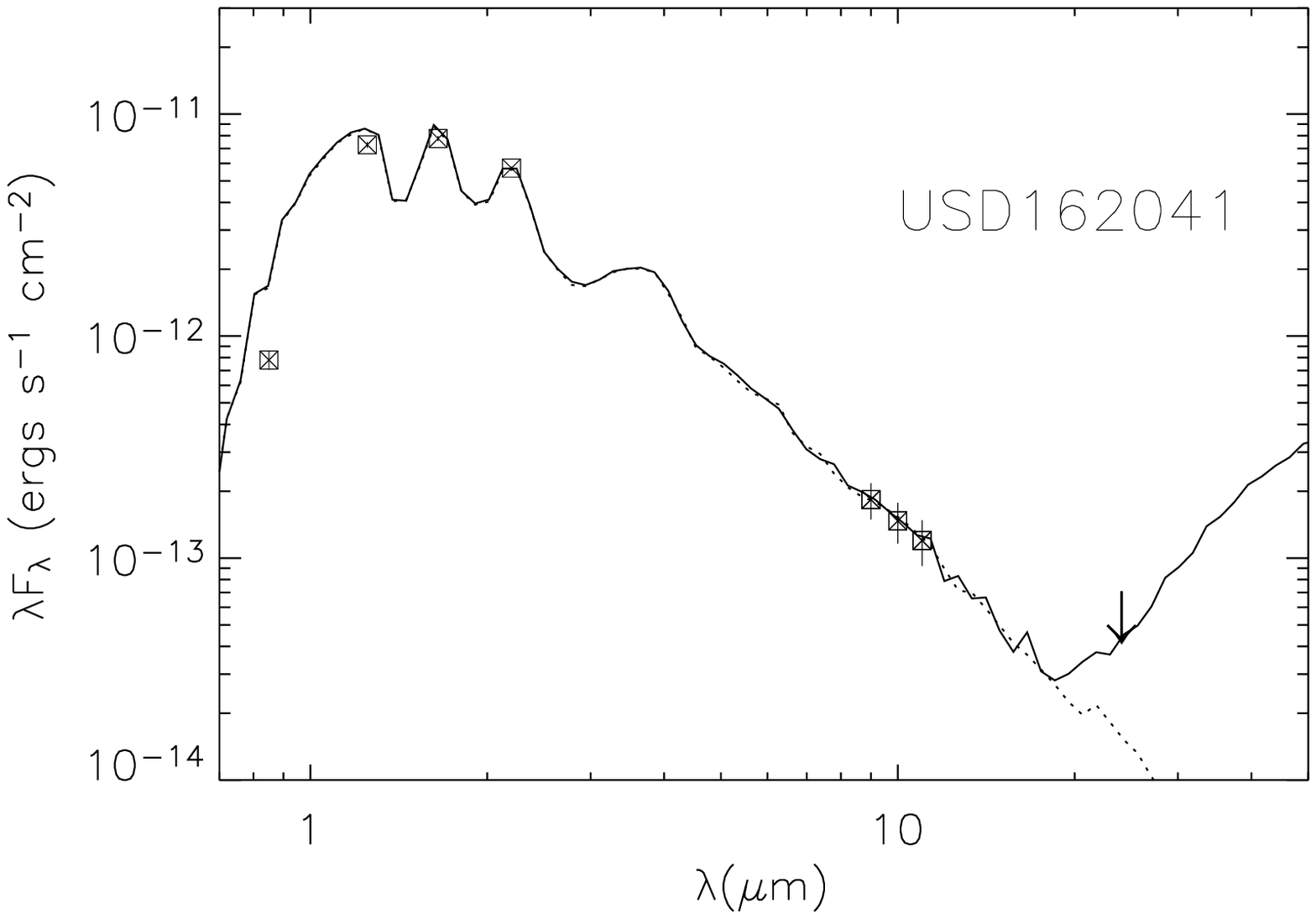}\\
\caption{Spectral energy distribution for objects without 24\,$\mu m$ MIR excess emission 
in comparison with models (dotted lines: photospheric flux, solid lines: photosphere + disk). 
The SEDs can be explained by photospheric emission plus disks with large inner holes 
(holesizes $\gtrsim 5$\,AU; see Sect. \ref{kenny}). \label{f3}}
\end{figure}

In Fig. \ref{f3} we show the results of our SED fitting for the objects with disk; Table \ref{modelparams} 
contains the star and disk parameters used to model each source. Each SED shows the data, the input 
atmosphere model spectrum (dotted line), our model SED using the parameters from Table \ref{modelparams} 
(solid line), and also the SED for a disk in hydrostatic equilibrium (dashed line, see 
\citet{2006ApJ...645.1498S} for a description of the hydrostatic disk models). For usd160958, we plot
additionally a model with disk radius of 5\,AU as dash-dotted line. All models are for a disk inclination 
angle $i=20^\circ$; none of our sources show the characteristic faint optical spectra indicative of 
highly inclined disks (see also Sect. \ref{objprop}).

As with our models for brown dwarf disks in Taurus, we immediately see that the highly flared
hydrostatic disk models, which assumes that dust and gas are well mixed, produce in all cases larger 
IR excesses than observed. Thus, some degree of dust settling has to be assumed to match 
the observed IRS and MIPS fluxes. The models assuming two dust components without hydrostatic 
equilibrium agree much better with the observations, consistent with a picture of dust grain growth 
and the settling of larger grains towards the disk midplane, leaving a population of small grains 
suspended at large heights producing the silicate feature seen in a few cases (Sect. \ref{sedgen} 
and \ref{gwen}). The best fit is typically achieved with a small amount of small grains or no small
grains at all, i.e. when the dust mass is dominated by the larger grains. The SED of usd160958 can 
be modeled adequately assuming a small disk radius of 5\,AU, again demonstrating that the mid-infrared 
is not very sensitive to the global disk radius (see Sect. \ref{bin} for further discussion of the 
disk-binarity connection).

\begin{figure}
\includegraphics[angle=0,width=13.8cm]{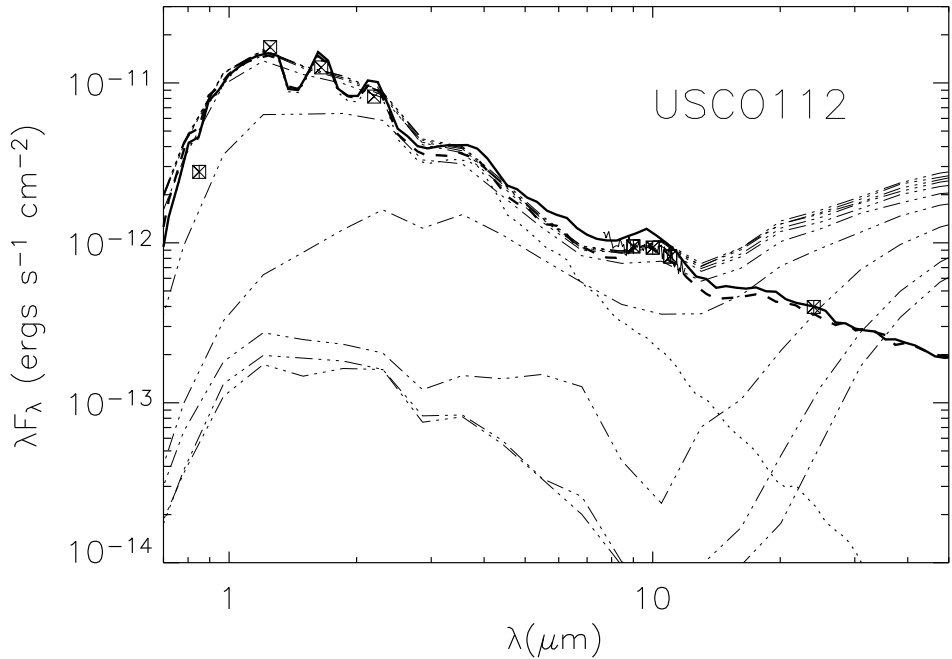}
\caption{Detailed SED modeling for the object usco112: The solid, thick line shows our best 
(non-hydrostatic) model already plotted in Fig. \ref{f2} (2nd panel), using small and large dust
grains. The dashed line is a best fit using the same model approach, but with small grains
only. Dash-dotted lines show hydrostatic models (with small grains only) viewed at ten different 
inclinations evenly spaced in cos(i) from i=0 to i=90. The photosphere is again shown as dotted line. 
\label{f8}}
\end{figure}

\begin{figure}
\includegraphics[angle=-270,width=13.8cm]{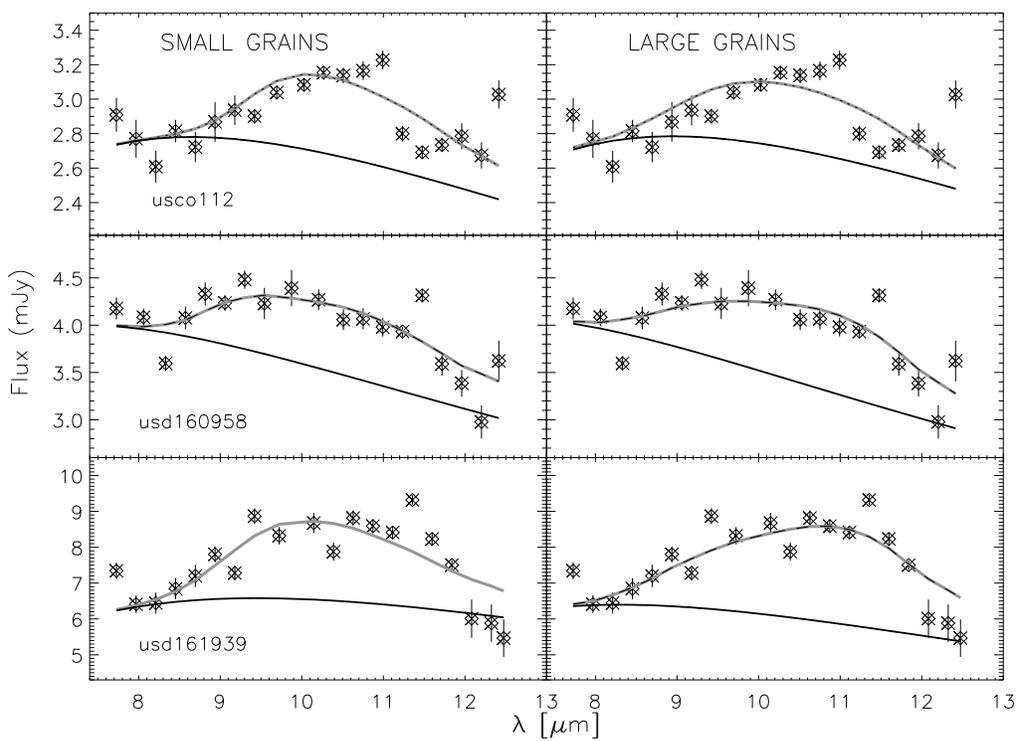}
\caption{IRS spectra for three objects with detection of dust emission features, together 
with uncertainties for the individual datapoints. Additional calibration uncertainties are 
$\sim 10$\%. The grey lines show a model fit to the feature using either small (left panel) 
or large grains (right panel), see Sect. \ref{gwen} for details of the modeling). As can be
seen, both grain species provide a good fit to the data. The black solid lines show the
continuum from the fit. \label{f4}}
\end{figure}

As described in \citet{2006ApJ...645.1498S}, disk structure, inclination, and grain sizes are 
degenerate parameters in these models. To confirm that dust settling is required to match the observed 
SEDs, we explored these degeneracies in more detail in Fig. \ref{f8}. For one exemplarily case, the 
object usco112, we plot our best fit using the non-hydrostatic model (solid line for large and small 
grains, dashed line for small grains only) and hydrostatic, highly flared models (with small grains only) 
for different inclinations (dash-dotted 
lines). As can be seen in this figure, for low and moderate inclinations the hydrostatic models give 
clearly too much flux at 24$\mu m$, which turns out to be strongest constraint for the disk 
geometry. Conversely, we can exclude high disk inclinations (i.e. close to edge-on geometry) based 
on the low extinction seen in the near-infrared photometry. Thus, some extent of dust settling 
and/or grain growth is required to explain the observed SED, no matter what value is chosen for 
the inclination. Fig. \ref{f8} also demonstrates that non-hydrostatic models with predominantly
large grains and with small grains only are both able to provide a reasonable match to the data,
because the large-scale mid-infrared SED is not particularly sensitive to the grain size. 

The results from the model fits can be directly compared to our previous findings for younger brown
dwarfs in Taurus, where we used the same models. In Taurus, however, we only analysed five out
of 20 objects with detection at 1.3\,mm; the sample is thus biased towards the most massive (and thus
maybe less evolved) disks. For all five Taurus objects a small amount of ISM-like dust is required to 
fit the model SED. The same is found in UpSco: For five of our disks, the best fit is obtained when 
we add small grains. The values for the flaring parameters $\beta$ are similar in both samples, but 
the scaling factors $h_0$ are much smaller (in absolute units) in all UpSco disks, leading to 
smaller scaleheights at a given radius. Thus, in our large sample in UpSco we do not see disks 
which are similarly strongly flared as the Taurus sources. This indicates a higher degree 
of dust settling in the UpSco objects when compared with younger brown dwarfs.

As there is a gap in our data coverage in the $3\,\mu$m to $8\,\mu$m region, we cannot constrain the size 
of any inner disk holes in sources showing IRS and MIPS excesses. Sources without MIR excess, however,
either have no disk or a disk with a inner hole, i.e. an optically thin region. To constrain the
inner hole sizes, we applied the same models to five objects without excess emission (usco75, usco104, 
usco130, usd1555605, usd162041). For four of these objects, the resulting model fits are shown in 
Fig. \ref{f3}. In all cases, a substantial inner hole is required to explain the lack of excess
emission, the minimum hole sizes are 5\,AU for usco130, usd155606, usd162041, 7.5\,AU for usco104, and
20\,AU for usco75. These estimates do not depend significantly on the choice of the grain size.

Taken together, the results from the SED modeling provide strong evidence for dust settling in 
UpSco brown dwarf disks. Hydrostatic, highly flared models cannot reproduce the observed SEDs. In 
addition, we find that $\lesssim 5-20$\,AU size (opacity) holes are required to fit the SEDs without 
disk excess.

\subsection{The silicate features}
\label{gwen}

The spectral region around 10\,$\mu m$ is valuable for deriving the
composition of the warm ($T>$200\,K) circumstellar dust. Indeed, it is here that 
the most common components of astronomical dust have spectral features. Using
these features, detailed analysis of the disk chemistry has been carried out 
previously for Herbig Ae/Be \citep[e.g.][]{2001A&A...375..950B} 
and T Tauri stars \citep[e.g.][]{2003A&A...409L..25M}. According to these studies,
the main dust species in circumstellar disks are olivines, pyroxenes and silica, 
either in amorphous or crystalline state. A similar modeling approach,
tailored to our faint targets, will be used in the following to analyse their IRS 
spectra. For a deeper discussion of the properties of astronomical dust and the 
mass absorption coefficients $\kappa$ we used in this paper, we refer to 
\citet{2005A&A...437..189V}. 

Given that the flux levels around 10$\,\mu m$ are low for most of our targets,
resulting in low signal-to-noise in the IRS spectra, we refrained from doing a 
detailed analysis of the dust composition, which requires high quality data. 
Instead, we tried to answer two questions using a more robust approach: a) Can we 
confirm the presence of silicate features in some of our targets (see Sect. 
\ref{sedgen})? b) Is there evidence for grain growth from these features?

Because changes in the dust composition only causes relatively subtle (for us 
undetectable) changes in the shape of the silicate feature, we simply adopted a 
dust composition, which contains only amorphous silicates, the most prominent 
dust species in many circumstellar disks. To check for evidence of grain growth,
we run the models with two different grain sizes, 0.1 and 1.5\,$\mu$m.
We represent the dust continuum by a blackbody with a temperature $T_{\mathrm{cont.}}$, 
and for simplicity, assumed that all the grains emitting at 10\,$\mu$m have the same 
temperature. Since we are modelling emission features, the region where the emission 
originates must be optically thin. Therefore, the flux of the emission feature can be 
represented by a linear combination of the mass absorption coefficents: 
$F_{\mathrm{emisson}}(\lambda) \propto B (T_{\mathrm{cont.}}) \sum_{i} a_{i} \kappa_{i}(\lambda)$, 
where $B (T_{\mathrm{cont.}})$ is the Planck function with a temperature 
$T_{\mathrm{cont.}}$, and $\kappa_{i}(\lambda)$ are the mass absorption coefficents 
and $a_{i}$ the multiplication factors for each individual species $i$ (in our case: 
only amorphous silicate). We used a reduced $\chi^{2}$ analysis to fit the spectra. For 
each object showing an indication of an emission feature, we obtain a best fit (lowest 
reduced $\chi^{2}$) for small and large grains, respectively.

In the following, we summarise the results for our 13 sources with confirmed disk
excess (see Sect. \ref{diskfreq}). For three objects, usco112, usd160958, usd161939, 
we detect a silicate feature, i.e. the spectrum can not be reproduced within the error bars
only by a blackbody dust continuum. This is in line with the results from visual inspection 
of the spectra done in Sect. \ref{sedgen}. For these three sources, we show the best fit for 
small and large grains in Fig. \ref{f4}. As can be seen in the figure, both small and large 
grains give an adequate fit to the spectra. The $\chi^2$ values for the best fit do not differ 
significantly for small and large grains, respectively. Thus, even when using binned data, the 
flux levels and thus the signal-to-noise in our sample are too low to distinguish between
different particle sizes from the IRS spectra.

For the remaining 10 objects with disk the dust emission features around 10\,$\mu m$ are either 
absent or too weak to be confidently detected in our spectra. The MIR SEDs from 8 to 
13\,$\mu m$ are mostly flat and featureless. Since the dust emission has its origin in 
relatively small ($<10\,\mu m$), warm grains in the disk atmosphere, this implies that to
some extent the dust has settled to the optically thick midplane, a process which is usually 
thought to be accompanied by grain growth \citep[e.g.][]{2004A&A...421.1075D}. Disk 
inclination might also contribute in some cases to weaken the silicate feature, although this is
not a major factor in our sample, because high inclinations (and thus close to edge-on disks) 
are excluded based on the low extinction (see Sect. \ref{objprop}).
 
In summary, we find silicate features for three out of 13 objects with disks. However, the
flux levels and thus the signal-to-noise ratios are too low to give meaningful constraints
on dust chemistry and particle sizes from the emission features. For the remaining 10 objects
with disk we do not detect an emission feature.

\section{Discussion: Brown dwarf disks at 5\,Myr}
\label{disc}

Both the purely empirical discussion of the MIR SEDs in Sect. \ref{sed} and the results 
from the modeling (Sect. \ref{kenny} and \ref{gwen}) provide a consistent picture of the disk
properties in our sample. Given our large sample of coeval targets, they can be used to give
constraints on evolutionary processes in brown dwarf disks. In the following, we will 
summarise the results from the previous sections and compare them with similar studies for younger
objects and more massive stars (Sect. \ref{dissipation} and \ref{chem}). Care has to be taken, though, 
when comparing results obtained at different wavelenghts, because longer wavelenghts probe the disk 
at larger radii. This is further complicated by the fact that when changing central object masses the 
same wavelength probes different regions in the disks. Nevertheless, as it turns out, the main conclusions 
are basically unaffected by these restrictions. Furthermore, we will include a discussion of the
accretion/disk connection (Sect. \ref{halpha}) and possible influences of binarity on disk evolution 
(Sect. \ref{bin}). 

\subsection{Disk dissipation}
\label{dissipation}

The main probe for disk dissipation timescales is the disk frequency. In Sect. \ref{diskfreq} 
we derive a disk frequency of $37\pm 9$\% for our sample of UpSco brown dwarfs, spectral types
$>M5$, based on the 24\,$\mu m$ MIPS fluxes. This is consistent with values of $50$\% derived 
previously based on smaller samples in the same region using shorter wavelength data  
\citep{2003AJ....126.1515J,2006astro.ph..8395B}. We can now compare these results with similar 
studies for more massive stars. Recently, \citet{2006ApJ...651L..49C} published a Spitzer disk survey 
for a large sample of stars spanning a wide range in spectral types in UpSco, thus coeval to our targets. 
For stars with spectral types K0-M5 (corresponding to masses 0.1-1$\,M_{\odot}$, complementary to our
targets), they report a disk frequency of $19 \pm 4$\%\footnote{We re-calculated the uncertainty based
on binomial statistics, to be consistent with the uncertainty estimate derived for our disk frequency.} 
derived from observations at 16$\,\mu m$ and a very similar value from observations at 8$\,\mu m$. This
is lower than our datapoint for brown dwarfs on a 1.8$\sigma$ confidence level.

Both samples, our brown dwarfs and the stars from \citet{2006ApJ...651L..49C},
are {\it unbiased} in the sense that the objects have been selected using criteria independent from
their disk properties. The only difficulty when comparing the two studies is the different wavelength used 
to probe the presence of inner disks. However, all objects for which we see strong excess at 24$\,\mu m$ do 
also show significant excess already at IRS wavelengths, i.e. between 8 and 12$\,\mu m$. In addition,
studies at wavelengths between 3 and 8$\,\mu m$ provide a similar disk frequency (see above). Conversely, 
the disk fraction derived by \citet{2006ApJ...651L..49C} does not change significantly when going from 
16$\,\mu m$ to 8$\,\mu m$. Apparently, the disk fraction is not a strong function of wavelength between 8 
and 24$\mu m$, regardless of object mass. Therefore, a comparison between the stellar disk frequency derived 
from 16$\,\mu m$ data and the substellar disk frequency obtained from 24$\,\mu m$ is valid. Moreover, both 
in the Carpenter et al. and in our sample the value for the disk frequency is fairly robust, because there is 
a clear gap in mid-infrared emission between objects with and without disks. Thus, this comparison suggests
that the disk fraction for 5\,Myr old brown dwarfs (spectral types $>$M5) is higher (on a 1.8$\sigma$ 
confidence level) than for coeval K0-M5 stars.

The disk frequency (and thus the average disk lifetime) in UpSco are therefore a function of 
stellar mass: As found by \citet{2006ApJ...651L..49C}, disks are less frequently observed around G/F stars 
compared with K0-M5 stars. Now we find that the disk frequency for objects with spectral types 
$>M5$ is even higher than for K0-M5. This indicates that the disk lifetime in UpSco, i.e. the age 
at which only very few objects ($\lesssim 10$\%) retain an inner disk, is anti-correlated with 
the object mass: $\lesssim 5$\,Myr for G/F stars, but larger than 5\,Myr for brown dwarfs.

To verify this result, it is useful to compare with IC348, a cluster with age of $\sim 3$\,Myr,
for which disk frequencies have been published recently: Based on IRAC data, \citet{2006AJ....131.1574L} 
derive disk fractions of $11\pm 8$\% for stars earlier than K6, $47\pm 12$\% for K6-M2 stars, and 
$28\pm 5$\% for M2-M6 stars. Using a similar method, \citet{2005ApJ...631L..69L} has previously found 
a disk frequency of $42\pm 13$\% for the brown dwarfs (spectral types $>$M6) in this cluster. Taken 
together, these results give evidence for a mass-dependent disk lifetime, however, it may not simply
be an increase towards lower masses as in UpSco. In pre-Spitzer studies, it was commonly found that 
brown dwarf disk lifetimes are not 'vastly different' \citep{2003AJ....126.1515J}
from those of T Tauri stars, i.e. for both object classes the disk frequency drops to values $<10$\%
after 5-10\,Myr \citep{2001ApJ...553L.153H}. While this conclusion still holds, Spitzer may now allow
us to determine disk lifetimes as a function of mass more accurately. The results from IC348 
\citep{2006AJ....131.1574L,2005ApJ...631L..69L} and UpSco \citep[this paper,][]{2006ApJ...651L..49C}, 
although not entirely consistent, are first steps in this direction. Future studies in other star 
forming regions will hopefully help to disentangle the effects of mass, environment, and age on 
the disk dissipation timescale.

Our data provide useful constraints for the duration of the inner disk clearing process. The 
brown dwarfs in UpSco clearly fall in two groups: About one third has significant MIR excess, whereas 
the MIR emission for the remaining objects is completely consistent with photospheric fluxes (see
Sect. \ref{diskfreq}). For the latter ones, we find that they either have no disks or disks with a
large inner hole having a radius of at least 5-20\,AU (see Sect. \ref{kenny}). "Inner hole" in 
this context means an "opacity hole", i.e. the clearing from small ($\lesssim 1$\,mm) and thus 
observable dust grains; this does not necessarily imply that the region is completely devoid of 
material. 

This clear dichotomy and the lack of intermediate objects indicate that the dust clearing in the
inner 5\,AU of the disks occurs rapidly. Assuming that our objects have an age spread
of 2\,Myr (see discussion in Sect. \ref{objprop}) and are evenly distributed in age,
the time scale for this process is shorter than 2\,Myr divided by the number of objects
in our sample. Thus, a rough upper limit for the timescale of inner dust clearing within
5\,AU is $\sim 10^5$ years. The rapid clearing process is also supported by the fact 
that disk frequencies do not strongly vary when derived from observations in different
wavelength regimes (i.e. from 3 to 24\,$\mu m$, see above). 

The consequence is a pronounced scarcity of so-called transition disks, with lack of emission at 
$\lambda < 10\,\mu m$, but excess at 24\,$\mu m$ and thus a large hole in the inner disk. 
A few of these transition disks have been found recently and they are believed to be 'caught in the 
act' of inner disk clearing \citep[e.g.][]{2002ApJ...568.1008C,2005ApJ...630L.185C}. From our sample, 
we can constrain the frequency of transition disks among brown dwarfs in UpSco to be less than
8\% (2$\sigma$ confidence limit). Since our wavelength coverage has a gap between 2.5 and
7.5\,$\mu m$, this does not include objects with hole radii of $\lesssim 1$\,AU. Therefore,
our result is consistent with transition disk frequencies found for very low mass objects 
in IC348 \citep{2006ApJ...643.1003M} as well as for stars in young clusters 
\citep[e.g.][]{2006ApJ...638..897S}. Based on the available data, we conclude that 
inner disk clearing timescales are not vastly different between stars and brown dwarfs.

It should be emphasised that the inner disk clearing timescale is not comparable to the
inner disk lifetime. Inner disk clearing appears to be a rapid process, which
may be triggered by photoevaporation \citep{2006MNRAS.369..229A} or formation of giant
planets. It interrupts the relatively quiet viscous evolution of the disk, produces large 
inner opacity holes, and may initiate the transformation of T Tauri like accretion disks to 
debris disks. From all what we know today, this scenario does probably also apply in the 
substellar regime.

\subsection{Dust settling and grain growth}
\label{chem}

Our results regarding dust evolution can be summarised as follows: Out of
13 definitive disk detections, 10 have a flat broad-band SED (measured in fluxes, 
within our uncertainties) from 8 to 24\,$\mu m$, and only three show a weak flux increase 
at 24\,$\mu m$. This is best explained by dust settling to the disk midplane, which
is usually thought to be accompanied by grain growth. The radiative transfer modeling 
in Sect. \ref{kenny} confirms this result; for all objects with disk, some degree of
dust settling is required to match the observed broad-band SED.

For a model-independent evaluation of the geometric properties of the disks, we 
calculated the flux ratio $F_{24/8} = F(24\mu m)/F(8 \mu m)$ for the brown dwarf disks 
in UpSco (13 objects) and for five brown dwarfs in Taurus, recently investigated in 
\citet{2006ApJ...645.1498S}. This flux ratio is a robust indicator for the flaring 
in the disk; it is expected to be around or below unity for flat disks and will exceed 
unity significantly for strongly flared disks \citep[e.g.][]{2002ApJ...567.1183W,2004ApJ...609L..33M}. 
From the observed SEDs, we obtain $F_{24/8} = 0.5 \ldots 1.6$ in UpSco and 
$F_{24/8} = 1.2 \ldots 3.8$ in Taurus. For three out of five objects in Taurus
$F_{24/8}$ exceeds the highest value in UpSco. Thus, our sample in UpSco lacks 
the highly flared disks typical for very young star forming regions, indicating
a larger degree of dust settling. This confirms the results from comparing the
outcomes of SED models for UpSco and Taurus discussed in Sect. \ref{kenny}. Again, 
it has to be emphasised that while our survey in UpSco is unbiased with respect to 
the disk properties, the small Taurus sample comprises only the objects with the most 
massive disks.

While the dust settling can be probed reasonably well with our data, it is more
difficult to evaluate the grain properties. For three objects in our sample, we 
find significant silicate emission features (see Sect. \ref{gwen}), indicating the 
presence of hot, small grains. Following \citet{2005Sci...310..834A} (see their Fig. 2), 
we determined parameters for the shape (continuum-subtracted flux ratio 
$F(11.3\mu m)/F(9.8 \mu m)$) and strength (peak flux above the continuum) for 
these three objects, using the continuum derived in Sect. 
\ref{gwen}. We obtain values of 0.5 to 1.5 ($\pm 0.2$) for the peak strength and 1.2 
to 1.6 ($\pm 0.05$) for the shape. These numbers are not strikingly different from 
what has been found for brown dwarfs in ChaI \citep{2005Sci...310..834A}. It has to be 
clarified that \citet{2005Sci...310..834A} have targeted objects known to be accreting and/or 
bright in the MIR. Thus, their sample is biased towards objects with strong MIR emission. 
Therefore, we basically compare the objects with the {\it strongest} silicate features in
UpSco and ChaI, and their dust properties (based on shape and strength of the emission
feature) appear to be rather similar. This is based on small samples and should therefore 
be substantiated as soon as more data is available. To take into account the
large number of objects {\it without} measurable silicate feature in UpSco, 
it would be particularly interesting to compare with a complete, unbiased brown 
dwarf IRS survey in a younger region like ChaI or Taurus, which is not available yet.

Neither the large-scale SED nor the silicate feature allow us to put much constraint on 
the coagulation of grains and their sizes. There are, however, two indirect arguments
for the occurence of grain growth in UpSco disks: a) the evidence for dust settling,
which is usually thought to happen as the grains coagulate and decouple from the gas, and
b) the absence of strong silicate emission in the majority of the UpSco disks and thus
the lack of hot, small dust grains. Based on these indirect arguments, the occurrence
of grain growth is ubiquitous in our sample.

The general picture of the dust evolution in brown dwarf disks by dust settling and grain 
growth is consistent with recent findings for stars. All these processes are now routinely 
observed with Spitzer in T Tauri disks \citep[e.g.][]{2005ApJ...634L.113M,2005ApJ...628L..65F,2005prpl.conf.8565B,2004ApJS..154..443F}.
It is not clear yet, however, to what extent the dust evolution is a function of object mass.
This can now be probed in UpSco by comparing the broad-band SEDs of stars and brown dwarfs. 
In Fig. \ref{f9}, we overplot the typical flux levels for K0-M5 stars with and without disks at 
8 and 16$\,\mu m$ as solid and dashed lines (taken from \citet{2006ApJ...651L..49C}). Although 
the wavelength coverage for stars and brown dwarfs is not identical, hampering a reliable comparison, 
it is clear that the mid-infrared fluxes of the brown dwarfs with disks clearly fall in the range 
defined by the sample of stars. The spread in the mid-infrared SEDs appears to be smaller in the 
brown dwarf regime by 0.5-1 order of magnitude, maybe indicating a more uniform evolutionary 
timeline for very low mass objects. But there is no clear evidence that the typical slope in the 
SED and thus the efficiency of the dust settling process in the inner disk is a strong function of 
mass. This is contrary to the results of \citet{2006AJ....131.1574L} in the somewhat younger cluster
IC348, where a larger degree of dust settling is seen for late M-type objects. To what extent 
this discrepancy is a result of age effects or environmental differences has to be clarified
in future work.

\subsection{Accretion signature from H$\alpha$ emission}
\label{halpha}

We have compiled H$\alpha$ 10\% width measurements for our targets from our own spectra and
the literature (see Table \ref{targets}). This is the largest sample of brown dwarfs for
which both MIR SEDs and H$\alpha$ measurements are available. H$\alpha$ emission is a useful 
indicator of disk accretion, and particulary the 10\% width has been found to be correlated with the 
accretion rate \citep{2004A&A...424..603N}. Typically, a 10\% width of $\sim 300$\,kms$^{-1}$
corresponds to accretion rates of $\sim 10^{-10}\,M_{\odot}$/year. A value of 200-270\,kms$^{-1}$
is usually adopted as a threshold between accretors and non-accretors, corresponding to
accretion rates below $\sim 10^{-11}\,M_{\odot}$/year 
\citep{2003ApJ...582.1109W,2003ApJ...592..282J,2005ApJ...626..498M}. It is important to emphasise 
that this is mostly an empirical limit and in rare cases even a non-accretor might exceed the
threshold because of fast rotation or strong magnetic activity. 

\begin{figure}
\includegraphics[angle=-90,width=13.8cm]{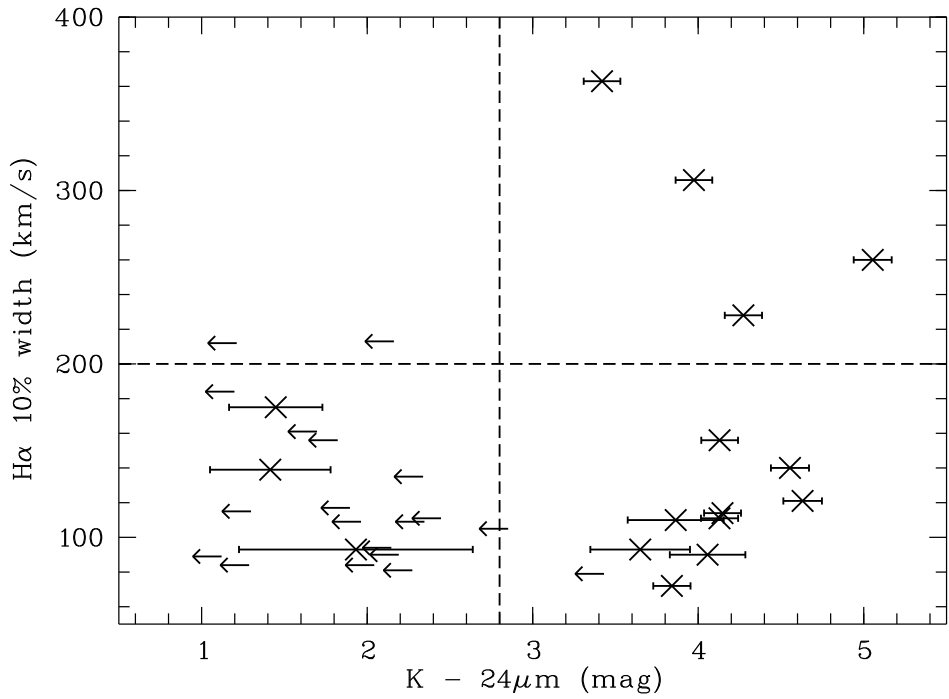}
\caption{H$\alpha$ 10\% width vs. K-band minus 24$\mu m$ colours for UpSco brown dwarfs. Upper limits 
correspond to 2$\sigma$ upper limits at 24$\mu m$. The dashed horizontal line separates accretors 
from non-accretors, while the dashed vertical line separates objects with infrared excess from those
without (see Sect. \ref{halpha}).  \label{f5}}
\end{figure}

To probe the connection between disks and accretion, we plot the H$\alpha$ 10\% widths vs. 
the $K - 24\,\mu m$ colour already used in Sect. \ref{diskfreq} (see Fig. \ref{f5}). The
four quadrants in this figure correspond to the expected positions for objects without disks
and without accretion (lower left), with disks but without accretion (lower right), with disks
{\it and} accretion (upper right), and without disks but with accretion (upper left). 
The left half of this figure is easy to understand: There are no objects without mid-infrared
excess and H$\alpha$ 10\% width clearly exceeding the 200\,kms$^{-1}$ limit. Thus, when an object 
has lost the disk or developed a inner opacity hole of at least 5\,AU, the accretion also drops 
to unmeasurable rates. Thus, once the inner 5\,AU of the disk is devoid of dust, the accretion
rate drops below $\sim 10^{-11}\,M_{\odot}$/year.

The right half of Fig. \ref{f5} is more puzzling. Surprisingly, only four out of 13 objects 
with clear evidence for the presence of an inner dust disk show a spectroscopic signature of 
accretion. The remaining nine sources with inner disks are located well below the threshold 
between accretors and non-accretors, so their accretion rates are probably significantly lower 
than $\sim 10^{-11}\,M_{\odot}$/year. Thus, although a dusty disk is still present in the inner
5\,AU around these objects, accretion has ceased or at least reduced to a trickle. This confirms
previous claims by \citet{2003AJ....126.1515J} and \citet{2006astro.ph..8395B} based on NIR/MIR 
observations of smaller samples.

It is not clear to what extent the lack of accretion in these nine objects with disk is an 
indication for depletion of gas in the disk, particularly because two of these nine have the 
most flared disks in our sample. Flaring requires the presence of small grains (see Sect. 
\ref{sedgen} and Sect. \ref{kenny}), and it is usually believed that due to dust/gas coupling 
small dust grains are a good tracer for the presence of gas \citep[e.g.][]{2001ApJ...553L.153H}. 
Thus, although accretion is not detectable in these seven objects, there is presumably still 
a (small?) amount of gas in their disks.

A plausible explanation for the lacking disk/accretion connection might be variable 
accretion. Spectroscopic signatures of accretion are known to be variable in accreting objects
\citep[e.g.]{1995AJ....109.2800J,2002ApJ...571..378A,2003A&A...409..169B}. Specifically
for young brown dwarfs, we found recently that accretion rate changes by 0.5-1 order of magnitude
and episodic phases of accretion and quiescence are not unusual 
\citep{2005ApJ...629L..41S,2006ApJ...638.1056S}. Assuming that the phases of quiescence are dominant 
in the latest stages of accretion, maybe due to the onset of inner disk clearing, it might be that 
most objects with disks are observed in an epoch of low (and thus unmeasurable) accretion level, 
although they still have an inner disk. Multi-epoch spectra are required to constrain the degree of 
accretion variation in these objects. In summary, based on our large sample we find that the inner 
disks of brown dwarfs can persist longer than measurable rates of continuous accretion.
 
\subsection{Binarity and disk evolution}
\label{bin}

Large parts of our sample have been observed with high spatial resolution, either with
HST or with AO imaging using NACO at VLT \citep{2006A&A...451..177B,2005ApJ...633..452K}. 
The main goal of these surveys was to search for companions around young brown dwarfs; the
observations are typically sensitive to substellar companions for separations 
between 5-10\,AU and several hundred AU. The results are summarised in Table \ref{targets}: 
Out of 29 objects that have been covered, 6 have been found to have a companion in the given 
separation range. These companions are unresolved in our Spitzer images and spectra. In 
the following, all separations are given as projected values, assuming a distance of 145\,pc 
for UpSco. Strictly speaking, these values should be considered to be lower limits to the 
physical separation.

Two of the objects with companions in our sample, usco66 and usco109, do not show any MIR excess, 
i.e. they have either lost their disk or the disk has a large inner hole. These are also the two
objects with the smallest separations between the two binary components, 10 and 5\,AU, respectively.
The four remaining binary brown dwarfs (usco55, usd160958, usd161833, usd161939) in the sample, with 
separations ranging from 12\,AU to 130\,AU, show clear signs for a dusty disk. 

The existence of companions inside 10\,AU for objects without MIR excess is in line with
the idea that these objects have either lost their disk or harbour a disk with a large 
inner hole, which would then be circumbinary. For the objects with inner disk {\it and} companion
the situation is more complex. The presence of a companion is usually thought to truncate
the individual disks of the components due to tidal forces, leaving two objects with small
disks plus potentially an outer circumbinary disk \citep[see][]{1994ApJ...421..651A}, which
however would be undetectable with our MIR data. For solar-mass binaries with mass ratios 
close to one, the tidal truncation radius is between 0.3 and 0.4 times the (physical) separation 
\citep{1999MNRAS.304..425A}. For the closest binaries with disk -- usd160958, usco55, and usd161939 
-- this might imply outer disk radii as low as 5-10\,AU (assuming that projected separation is
comparable to the physical separation). This scenario of disk truncation due to close companions 
is not inconsistent with our MIR SEDs: As we have shown in Sect. \ref{kenny} for the case of usd160958 
(which has a 12\,AU companion), it is possible to have a small disk within the truncation radius that 
gives enough dust emission to explain the measured flux levels in the MIR (see Fig. \ref{f2}, 7th panel). 
Thus, it appears plausible that these three objects feature small disks and companions. 

All three objects with increasing flux levels in the MIR (and thus flaring) have 
a companion, at projected separations of 18, 26, and 130\,AU, as already pointed out in Sect. 
\ref{sedgen}. At face value, this might be a hint for enhanced flaring in the inner disk due
to a companion, e.g. by increased disk heating or turbulence. The literature, however, does not
provide support for this scenario: Large-scale studies for T Tauri stars carried out by 
\citet{2006ApJS..165..568F} and \citet{2006ApJ...636..932M} find no correlation of MIR properties 
and the presence of companions. Thus, our finding of three objects with flared disks and
companions might well be result of coincidence and should be treated with caution.

\section{Summary}
\label{conc}

We have carried out a Spitzer survey of 35 brown dwarfs in the Upper Scorpius star forming region
with an age of $\sim 5$\,Myr. Combined with literature data from spectroscopy and high-resolution 
imaging, the brown dwarfs in UpSco now comprise a large sample of substellar objects, for which 
information about disks, accretion, and multiplicity is available. In the following, we list our 
most important results:

\begin{enumerate}
\item{Out of 35 objects, 13 show clear excess emission at 24$\,\mu m$, indicating emission 
from a dusty circum-sub-stellar disk. This corresponds to a disk frequency of $37\pm 9$\%. The 
disk frequency is on a 1.8$\sigma$ confidence level higher in our sample than for K0-M5 stars 
in the same region \citep[$19\pm 4$\%][]{2006ApJ...651L..49C}, suggesting a mass-dependency in the 
disk lifetime.}
\item{The objects without emission either do not have a disk or they harbour a disk with large
inner holes (radii $>5$\,AU). The clear distinction between objects with/without disks and the 
lack of objects in transition between the two subsamples is a clear sign of rapid inner disk
clearing, probably on timescales of $\lesssim 10^5$ years.}
\item{The midinfrared SEDs of the disks appear uniform: Most of them are flat between 8 and
24$\,\mu m$ with flux levels of a few mJy. Only three objects show slighly enhanced flux levels
at 24$\,\mu m$, indicative of flaring. Hydrostatic disk models where dust and gas are well-mixed
are not able to fit the data. However, assuming a two component dust composition where 
large grains have smaller scaleheights than ISM-like small grains provides good agreement 
between model and observations. This is well-explained with a scenario in which grain growth 
leads to larger grains that settle to the disk midplane.}
\item{In most disks, there is no clear indication for dust emission features around 10$\,\mu m$,
arguing for a lack of hot, small grains. This confirms that the disks are affected by dust settling 
and grain growth.}
\item{Brown dwarf disks in UpSco (5\,Myr) show a larger degree of dust settling than in the younger
Taurus star forming region. In UpSco itself, we do not see a significant difference between stars
and brown dwarfs in the broad-band SEDs, indicating that dust settling is not a strong function
of object mass.}
\item{Taken together, our survey finds disks at an advanced stage: A large fraction of the objects
have already cleared out inner parts of the disk. Evolutionary processes like dust settling and grain 
growth are ubiquitous in brown dwarf disks at 5\,Myr. Since these processes are prerequisites for 
the formation of larger bodies in the disk, our results suggest possible planetesimal growth and 
planet formation in brown dwarf disks.}
\item{Out of 13 objects with disks, only four appear to accrete at more than 
$\sim 10^{-11}\,M_{\odot}$/year. Thus, dusty disks can persist although continuous 
accretion has stopped or at least reduced to unmeasurable levels.}
\item{Four objects with disks also have a companion; in three cases the (projected) separations are 
smaller than 30\,AU. This can be explained with the existence of a small disk (radius 5-10\,AU) around
the primary, truncated by the companion.}
\end{enumerate}

\acknowledgments
This work is based on observations made with the Spitzer Space Telescope in the framework
of the GO program \#20435 (PI: Ray Jayawardhana) and the GTO program \#248 (PI: Subhanjoy
Mohanty). We thank Subhanjoy Mohanty for unbureaucratic data exchance and stimulating discussions.
The help of Duy Cuong Nguyen, who was the observer at Magellan, is gratefully acknowledged. 
Fruitful discussions with the star formation group in Toronto, in particular with Yanqin 
Wu, have helped to improve the paper. We thank the referee for a thorough and constructive 
report. This project makes use of data products from the Two Micron All Sky Survey, 
and was supported by an NSERC grant and University of Toronto research funds to Ray Jayawardhana.
Gwendolyn Meeus acknowledges financial support by the Deutsche Forschungsgemeinschaft (DFG) 
under grant ME2061/3-1.

Facilities: \facility{Spitzer,Magellan,ESO/VLT}

\clearpage

\begin{deluxetable}{llllcccccccc}
\tabletypesize{\scriptsize}
\tablecaption{Object parameters and mid-infrared fluxes. \label{targets}}
\tablewidth{0pt}
\tablehead{
\colhead{Name} & \colhead{$\alpha$(J2000)} & \colhead{$\delta$(J2000)} &\colhead{SpT} & \colhead{$T_{\mathrm{eff}}$} 
& \colhead{H$\alpha$ 10\%} & Bin.\tablenotemark{d} & \colhead{F(9$\mu m$)} & \colhead{F(10$\mu m$)} 
& \colhead{F(11$\mu m$)} & \colhead{IRS err.} & \colhead{F(24$\mu m$)\tablenotemark{a}}\\
\colhead{} & \colhead{} & \colhead{} & \colhead{} & \colhead{(K)} & \colhead{(kms$^{-1}$)} & \colhead{} &
\colhead{(mJy)} & \colhead{(mJy)} & \colhead{(mJy)} & \colhead{(mJy)} & \colhead{(mJy)}}
\tablecolumns{12}
\startdata
usco53    & 16 00 26.3 & -22 59 40.4  & M5   & 3100 &  175\tablenotemark{c}  & NaN & 2.36 &	2.03 &    1.73  &   0.28 &  $0.79\pm 0.20$		\\  
usco55    & 16 02 45.6 & -23 04 49.8  & M5.5 & 3050 &  114\tablenotemark{c}  & 18  & 6.40 &	6.49 &    6.43  &   0.29 &  $8.16\pm 0.82$\tablenotemark{b}\\
usco66    & 16 01 49.7 & -23 51 07.4  & M6   & 3000 &  115\tablenotemark{c}  & 10  & 1.33 &	1.38 &    0.80  &   0.27 &  $<0.4$	 \\
usco67    & 15 59 25.9 & -23 05 08.1  & M5.5 & 3050 &  139\tablenotemark{c}  & S  & 1.69 &     1.49 &	 1.16  &   0.24 &  $0.62\pm 0.20$\\
usco75    & 16 00 30.2 & -23 34 44.7  & M6   & 3000 &  212\tablenotemark{c}  & S  & 1.10 &     0.92 &	 0.75  &   0.20 &  $<0.4$	\\
usco100   & 16 02 04.1 & -20 50 41.5  & M7   & 2850 &  184\tablenotemark{c}  & S  & 0.98 &     0.93 &	 0.83  &   0.29 &  $<0.4$	\\
usco104   & 15 57 12.7 & -23 43 45.3  & M5   & 3100 &  109\tablenotemark{c}  & NaN & 0.64 &	0.58 &    0.49  &   0.17 &  $<0.4$	 \\
usco109   & 16 01 19.1 & -23 06 38.6  & M6   & 3000 &	84\tablenotemark{c}  & 5  & 0.71 &     0.64 &	 0.46  &   0.33 &  $<0.4$	\\
usco112   & 16 00 26.6 & -20 56 32.0  & M5.5 & 3050 &  111\tablenotemark{c}  & S  & 2.85 &     3.11 &	 3.01  &   0.13 &  $3.18\pm 0.32$\tablenotemark{b}\\
usco128   & 15 59 11.2 & -23 37 59.0  & M7   & 2850 &  121\tablenotemark{c}  & S  & 2.11 &     2.20 &	 2.23  &   0.18 &  $2.65\pm 0.27$\tablenotemark{b}\\
usco130   & 15 59 43.6 & -20 14 38.1  & M7   & 2850 &  111\tablenotemark{c}  & S  & 0.64 &     0.44 &	 0.41  &   0.13 &  $<0.4$\tablenotemark{e}	\\
usco131   & 16 00 19.5 & -22 56 28.9  & M6.5 & 2900 &  105\tablenotemark{c}  & S  & 0.29 &     0.43 &	 0.37  &   0.24 &  $<0.4$	\\
usd155556 & 15 55 56.0 & -20 45 18.5  & M6.5 & 2900 &  140		    & S   & 5.19 &     4.82 &	   4.90  &   0.24 &  $4.42\pm 0.44$\tablenotemark{b}\\
usd155601 & 15 56 01.0 & -23 38 08.1  & M6.5 & 2900 &  109		    & S   & 2.98 &     2.74 &	   2.68  &   0.21 &  $3.56\pm 0.36$\tablenotemark{b}\\
usd155605 & 15 56 05.0 & -21 06 46.4  & M7   & 2850 &  156		    & S   & 0.48 &     0.51 &	   0.38  &   0.29 &  $<0.4$	  \\
usd160440 & 16 04 40.8 & -19 36 52.8  & M6.5 & 2900 &  117		    & S   & 0.64 &     0.56 &	   0.64  &   0.22 &  $<0.4$	  \\
usd160455 & 16 04 55.8 & -23 07 43.8  & M6.5 & 2900 &  213		    & S   & 0.86 &     0.73 &	   0.22  &   0.26 &  $<0.4$	  \\
usd160514 & 16 05 14.0 & -24 06 52.6  & M6   & 3000 &	84		    & S & 1.36 &    1.29 &	  0.93  &   0.22 &  $<0.4$	  \\
usd160603 & 16 06 03.9 & -20 56 44.6  & M7.5 & 2650 &  306\tablenotemark{c}  & S  & 3.41 &     3.49 &	    3.60  &   0.18 &  $2.84\pm 0.28$\tablenotemark{b}\\
usd160951 & 16 09 51.1 & -27 22 42.2  & M6   & 3000 &  156		    & S   & 1.14 &     1.08 &	   0.68  &   0.29 &  $<0.4$	  \\
usd160958 & 16 09 58.5 & -23 45 18.6  & M6.5 & 2900 &  363		    & 12 & 4.32 &    4.23 &	   4.03  &   0.24 &  $3.73\pm 0.37$\tablenotemark{b}\\
usd161005 & 16 10 05.4 & -19 19 36.0  & M7   & 2850 &  228		    & S   & 2.81 &     2.23 &	   1.99  &   0.51 &  $3.04\pm 0.30$\tablenotemark{b}\\
usd161006 & 16 10 06.0 & -21 27 44.6  & M8.5 & 2500 &  110\tablenotemark{c}  & NaN & 0.91 &	1.52 &      1.62  &   0.54 &  $0.78\pm 0.20$\tablenotemark{b}\\
usd161007 & 16 10 07.5 & -18 10 56.4  & M6   & 3000 &	89		    & S   & 2.27 &     1.79 &	    1.01  &   0.60 &  $<0.4$	   \\
usd161103 & 16 11 03.6 & -24 26 42.9  & M9   & 2400 &	90\tablenotemark{c}  & NaN & 1.61 &	1.61 &      1.31  &   0.34 &  $0.99\pm 0.20$\tablenotemark{b}\\
usd161452 & 16 14 52.6 & -20 17 13.2  & M9   & 2400 &	79\tablenotemark{c}  & NaN & 0.15 &	0.15 &      -0.02 &   0.11 &  $<0.4$	   \\
usd161632 & 16 16 32.2 & -22 05 20.2  & M6   & 3000 &	93		    & S   & 1.83 &     1.68 &	   0.95  &   0.52 &  $0.35\pm 0.20$  \\
usd161833 & 16 18 33.2 & -25 17 50.4  & M6   & 3000 &	72		    & 130 & 5.35 & 4.37 &   4.16  &   0.33 &  $7.47\pm 0.75$\tablenotemark{b}\\
usd161840 & 16 18 40.8 & -22 09 48.1  & M7   & 2850 &	90		    & S   & 0.70 &     0.29 &	   0.51  &   0.51 &  $<0.4$	  \\
usd161903 & 16 19 03.4 & -23 44 08.8  & M6.5 & 2900 &	94		    & S   & 0.76 &     0.86 &	   0.78  &   0.28 &  $<0.4$	  \\
usd161916 & 16 19 16.5 & -23 47 22.9  & M8   & 2600 &	93\tablenotemark{c}  & NaN & 1.06 &	0.88 &      0.65  &   0.45 &  0.5-1\tablenotemark{b}\\
usd161926 & 16 19 26.4 & -24 12 44.5  & M6   & 3000 &  161		    & S   & 0.80 &     0.84 &	    0.16  &   0.27 &  $<0.4$	   \\
usd161929 & 16 19 29.9 & -24 40 47.1  & M8   & 2600 &  135\tablenotemark{c}  & S  & 0.49 &     0.92 &	    0.41  &   0.29 &  $<0.4$	   \\
usd161939 & 16 19 39.8 & -21 45 35.1  & M7   & 2750 &  260		    & 26  & 7.60 &     8.53 &	    8.78  &   0.52 &  $10.7\pm 1.1$\tablenotemark{b}\\
usd162041 & 16 20 41.5 & -24 25 49.0  & M7.5 & 2650 &	81\tablenotemark{c}  & S  & 0.55 &     0.49 &	    0.44  &   0.41 &  $<0.4$	   \\
\enddata
\tablenotetext{a}{Upper limits are based on 2$\sigma$ flux upper limits. Errors correspond to
1$\sigma$ flux uncertainties.}
\tablenotetext{b}{Safe disk detections according to the analysis in Sect. \ref{diskfreq}.}
\tablenotetext{c}{From \citet{2002ApJ...578L.141J} and \citet{2005ApJ...626..498M}.}
\tablenotetext{d}{Binarity: If a companion has been found, the projected separation in AU is given, 
assuming a distance of 145\,pc. A 'S' ('single') indicates that the object does not have a companion at 
separations $>5$\,AU. Objects with 'NaN' have not been observed yet with high spatial resolution 
\citep[see][]{2006A&A...451..177B,2005ApJ...633..452K}.}
\tablenotetext{e}{Also called usd160019, see \citet{2004AJ....127..449M}.}
\end{deluxetable}

\clearpage

\begin{deluxetable}{lccccccccc}
\tabletypesize{\scriptsize}
\tablecaption{Parameters from the radiative transfer modeling of the overall SED 
(see Sect. \ref{kenny})
\label{modelparams}}
\tablewidth{0pt}
\tablehead{
\colhead{Full name} & \colhead{$T_\star$} & \colhead{$R_\star$} & 
\colhead{$M_\star$} & \colhead{$R_\mathrm{out}$} &
\colhead{$h_0^{\rm big}$} & \colhead{$h_0^{\rm ISM}$} & \colhead{$f_{\rm ISM}$} & 
\colhead{$\beta$}  \\
\colhead{ } & \colhead{(K)} & \colhead{($R_\odot$)} & \colhead{($M_\odot$)} & \colhead{(AU)} &
\colhead{($R_\star$)} & \colhead{($R_\star$)} & \colhead{ } & \colhead{ } 
& \colhead{ }}
\tablecolumns{6}
\startdata
usco55      & 2800  & 0.65   & 0.26    & 100 &  0.015 & \dots  & 0.0 & 1.15  \\
usco112     & 2850  & 0.40   & 0.10    & 100 &  0.01 & 0.02  & 0.03 & 1.1   \\
usco128     & 2600  & 0.30   & 0.01    & 100 &  0.0375 & \dots  & 0.0 & 1.1  \\
usd155556   & 2900  & 0.40   & 0.075   & 100 &  0.06 & \dots  & 0.0 & 1.0   \\
usd155601   & 2900  & 0.35   & 0.075   & 100 &  0.03 & \dots  & 0.0 & 1.1   \\
usd160603   & 2900  & 0.35   & 0.03    & 100 &  0.05 & \dots  & 0.0 & 1.0   \\
usd160958   & 2900  & 0.55   & 0.075   & 100 &  0.01 & \dots  & 0.0 & 1.1   \\
            & 2900  & 0.55   & 0.075   & 5 &   0.007 & \dots  & 0.0 & 1.1  \\
usd161005   & 2850  & 0.33   & 0.04    & 100 & 0.025 & 0.03   & 0.03 & 1.1 \\	    
usd161006   & 2500  & 0.26   & 0.02    & 100 &  0.05 & 0.06  & 0.03 & 1.0  \\
usd161103   & 2500  & 0.26   & 0.02    & 100 &  0.05 & 0.06  & 0.03 & 1.0   \\
usd161833   & 3000  & 0.65   & 0.10    & 100 &  0.0055 & \dots  & 0.0 & 1.2  \\
usd161916   & 2600  & 0.25   & 0.03    & 100 &  0.015 & 0.025  & 0.03 & 1.1 \\
usd161939   & 2750  & 0.50   & 0.04    & 100 &  0.03 & 0.04  & 0.03 & 1.15   \\
\enddata
\end{deluxetable}

\clearpage

\clearpage

\clearpage

\clearpage

\clearpage

\clearpage

\clearpage

\end{document}